\documentclass[journal]{IEEEtran}
\usepackage{xcolor}
\usepackage{amsmath}
\usepackage{amsfonts}
\usepackage{amssymb}
\usepackage{graphicx}
\usepackage{subfigure}
\usepackage{array}
    \newcolumntype{C}[1]{>{\centering\let\newline\\\arraybackslash\hspace{0pt}}m{#1}}
    \newcolumntype{L}[1]{>{\raggedright\let\newline\\\arraybackslash\hspace{0pt}}m{#1}}
\usepackage{multirow}
\usepackage{ulem} 
\usepackage{cite}

\newcommand{\vect}[1]{\mathbf{#1}}
\newcommand{\dir}[1]{\hat{\vect{#1}}}

\newcommand{\kzup}{{k_z^\uparrow}}
\newcommand{\kzuppq}{k_{z,pq}^\uparrow}

\newcommand{\Zje}{Z^{J\rightarrow E}}

\begin{document}

\title{Efficient Tracking of Dispersion Surfaces for Printed Structures using the Method of Moments}

\author{Denis~Tihon,~\IEEEmembership{Member,~IEEE,}
        Shambhu~Nath~Jha,~\IEEEmembership{Senior Member,~IEEE,}
        Modeste~Bodehou,~\IEEEmembership{Member,~IEEE,}
        and~Christophe~Craeye,~\IEEEmembership{Senior Member,~IEEE}
\thanks{D. Tihon, M. Bodehou and C. Craeye are with the ICTEAM institute, Université catholique de Louvain (UCLouvain), Louvain-la-Neuve, Belgium. S. N. Jha is an Electromagnetics Specialist with Thales Belgium, Tubize and has performed this research work in the capacity of a Scientific Collaborator at Université catholique de Louvain (UCLouvain). This research contribution belongs to UCLouvain.}
\thanks{D. Tihon and M. Bodehou are postdoctoral reasearchers of the Fonds de la Recherche Scientifique - FNRS.}
\vspace{-0.9cm}
}

\markboth{IEEE Transactions on Antennas and Propagation,~Vol.~XX, No.~X, July~2023}%
{Shell \MakeLowercase{\textit{et al.}}: Bare Demo of IEEEtran.cls for Journals}

\maketitle

\begin{abstract}
The dispersion surfaces of printed periodic structures in layered media are efficiently computed using a full-wave method based on the periodic Method of Moments (MoM). The geometry of the dispersion surface is estimated after mapping the determinant of the periodic MoM impedance matrix over a range of frequencies and impressed phase shifts. For lossless periodic structures in the long-wavelength regime, such as lossless metasurfaces, a tracking algorithm is proposed to represent the dispersion surface as a superposition of parameterized iso-frequency curves. The mapping process of the determinant is accelerated using a specialized interpolation technique with respect to the frequency and impressed phase shifts. The algorithm combines a fast evaluation of the rapidly varying part of the periodic impedance matrix and the interpolation of the computationally intensive but slowly varying remainder. The mapping is further accelerated through the use of Macro basis functions (MBFs). The method has been first tested on lossless metasurface-type structures and validated using the commercial software CST. The specialized technique enables a drastic reduction of the number of periodic impedance matrices that needs to be explicitly computed. In the two examples considered, only 12 matrices are required to cover any phase shift and a frequency band larger than one octave. An important advantage of the proposed method is that it does not entail any approximation, so that it can be used for lossy structure and leaky waves, as demonstrated through two additional examples.
{\color{blue} Accepted version. \copyright 2023 IEEE.  Personal use of this material is permitted.  Permission from IEEE must be obtained for all other uses, in any current or future media, including reprinting/republishing this material for advertising or promotional purposes, creating new collective works, for resale or redistribution to servers or lists, or reuse of any copyrighted component of this work in other works.}
\end{abstract}

\begin{IEEEkeywords}
Periodic structures, dispersion curves, iso-frequency contours, fast tracking algorithm, interpolation, macro basis functions, metasurface.
\end{IEEEkeywords}

\IEEEpeerreviewmaketitle

\section{Introduction}
\IEEEPARstart{E}{lectromagnetic} fields in two-dimensional periodic structures have been studied for many different applications: phased arrays, photonic crystals, frequency selective surfaces, reflect-and-transmitarrays, metamaterials and metasurfaces. In this context, over the past four decades, the full-wave numerical analysis of such problems has been an intense subject of research \cite{Numerical_methods}. The study of infinitely periodic structures is generally simplified by studying a single unit cell of the structure and imposing the periodicity of fields and currents within a linear phase shift (hereafter referred to as \textit{quasi-periodicity}). From such analysis, global properties of the structure can be computed, such as the active impedances and reflection coefficients (both dependent on incidence or scan angle) or an homogenized equivalent sheet impedances \cite{Sheet_imped, chapter_maci}. A common way to characterize fields in periodic structures corresponds to dispersion surfaces, i.e. the 2D locus of points for which a homogeneous solution to Maxwell's equations exists in the 3D space of frequencies and impressed phase shifts, denoted respectively as $f$, $\phi_x$ and $\phi_y$. Many important physical quantities can be extracted from such surfaces, such as the presence of band-gaps, the Poynting vector direction, the phase and group velocities and the dispersion of electromagnetic pulses in the material \cite{D25}, with important applications in the design of new planar devices \cite{D27, D29}. 
Note that, in the presence of material or radiation losses, the frequencies or wave vectors can take complex values \cite{D14, D26}.

The methods used to compute the dispersion curves strongly depend on the geometry studied and the hypotheses made. If the fields entering into and escaping from the unit cell can be modelled using a finite number of degrees of freedom, the problem is mathematically simpler. First, the response of the unit cell to any external excitation is computed using a full-wave solver. Then, combining the response of the unit cell to the quasi-periodic boundary conditions, the phase shifts for which a homogeneous solution to Maxwell's equations exists, can be obtained by solving a linear eigenvalue problem. This formulation naturally applies to periodic structures whose unit cells have a finite extent, as is the case for 3D (\cite{D17, D22, D37,D1bis}), 2D (\cite{D19, D20, D24, D40}) or 1D periodic structures in 3D, 2D or 1D space, respectively, or for 1D and 2D periodic structures enclosed in a waveguide \cite{D29, D15, D16}. 
The method can also be applied to open geometries (i.e. infinite-extent unit cells) by artificially reducing the size of the unit cells using Perfectly Matched Layers (PML) \cite{D11, D23} or truncating the infinite set of modes \cite{D17, D16}. However, as highlighted in \cite{D16}, the selection of modes is a critical operation that may produce significant errors if not done properly. Moreover, the use of PMLs can degrade the results through the appearance of spurious reflections \cite{D11} or spurious guided modes \cite{D23}. Last, as illustrated in Section \ref{sec:cross_results}, these methods do not handle rigorously the continuous spectrum associated with radiation into unbounded ragilons (or ``space waves") \cite{space_waves_1, space_waves_2, space_waves_3}, potentially leading to implicit approximation \cite{D16}.

When dealing with truly open geometries, the computation of the dispersion curves is a non-linear problem. First, a function of the frequency and phase shifts is chosen, whose value vanishes (resp. diverges) in the presence of a homogeneous solution to Maxwell's equations (see e.g. \cite{D4bis}). Then, the dispersion curves are obtained by searching for the zeros (resp. poles) of the function in the $(f,\phi_x,\phi_y)$ space. This search can be carried out using iterative techniques or more involved methods based on the theory of complex analysis \cite{D5, D13}. In both cases, the function needs to be evaluated for many different phase shifts and frequencies. Using full-wave methods, it means that a new simulation is required at each iteration if no acceleration technique is used, leading to prohibitive computation times \cite{D36, D38}. To circumvent this problem, semi-analytical models have been proposed as a trade-off between the versatility of the full-wave modelling and the rapidity of analytical models. For example, the authors of \cite{D35, D31, D28, D34} proposed a semi-analytical model for the case of sub-wavelength printed patches on a sufficiently thick grounded substrate. The slowly varying response of the patches is modelled using an equivalent sheet impedance. The sheet impedance is estimated using full wave simulations and then interpolated for different phase shifts and frequencies. The impact of the substrate is added afterwards using a transmission line formalism. Using this method in combination with a few additional hypotheses, an analytical description of the dispersion curves can be obtained \cite{D32}. 

In this paper, we propose an accelerated technique for the full-wave computation of the dispersion surfaces of open 2D periodic metallizations on a stratified medium using the Method of Moments (MoM). The dispersion curves correspond to zeros of the determinant of the MoM impedance matrix. The acceleration relies on a specialized interpolation technique enabling a rapid sweep over frequency and phase shifts. ``Specialized" here means that a limited number of potentially singular terms is first removed, such that the interpolation needs to take place only over strictly smooth functions. The potentially singular terms are reintroduced at the end on the fine grid.

To the authors' best knowledge, it is the first time that a specialized interpolation method is proposed to handle both the impressed phase shifts and frequency. Several methods have been proposed to interpolate the periodic impedance matrix vs. frequency. In \cite{D1}, Hermite polynomials are used. To further accelerate the interpolation, the size of the matrix is decreased using an order reduction technique. In \cite{D8}, the authors propose to add a term to the polynomial expansion that is inversely proportional to frequency. In \cite{D9, D18, D4}, additional terms are added to account for the low-order Floquet modes, whose contribution can vary rapidly with frequency due to guided modes in the substrate. Last, the authors of \cite{D6} propose to compute analytically the contribution of the low-order Floquet modes to reduce the number of reference matrices needed to compute the coefficients of the expansion. The method proposed in this paper is an extension of the author's conference communications \cite{D2} and \cite{D3}, with an extension to interpolation versus frequency, reduced-order modeling of currents on the patches (see below), semi-automatic tracking of the iso-frequency contours and inclusion of radiation and substrate losses

After the rapid interpolation of the periodic impedance matrix, its determinant is evaluated. The metallizations are modelled using the standard Rao-Wilton-Glisson~\cite{RAO82} basis functions. Compared with specialized basis functions~\cite{Reviewers_3}, RWG have the advantage of being simple, standard and versatile, at the expense of a less compact current representation. Given the possibly large number of unknowns required to accurately model the patches, the time required to compute the determinant of the MoM matrix over a large number of points can be prohibitive. To reduce it, the currents on the patches are limited to a smaller subspace through the use of Macro Basis Functions (MBFs) \cite{D7bis}, leading to smaller impedance matrices. Another interpolation model is applied to the MBFs to directly interpolate the reduced impedance matrix. Using the second model, the logarithm of the determinant of the impedance matrix is rapidly evaluated on a regular grid of phase shifts and frequencies. 

In addition to the fast mapping strategy, we propose a tracking procedure to extract the dispersion curves from the determinant maps for lossless structures in the long-wavelength regime. 
For relatively low frequencies, i.e. periods smaller than a quarter wavelength (the metasurface range being rather from $10^{-1}$ to $2.10^{-1}$ wavelength), one can make the assumption that the iso-frequency contours will be closed. Under that assumption, the contours are described by an angle-dependent radius, formulated as a Fourier series. The tracking procedure establishes a harmonic model by giving weights to each pixel of MoM matrix determinant map versus phase shifts. An iterative
refinement of the model, particularly useful in presence of multiple iso-frequency contours, is also presented.
The tracking method is semi-automatic, since it involves a few parameters (e.g. relevance thresholds of pixels) that can depend on the type of patch analyzed.

It should be noted that, in addition to its relatively low computation time, the proposed method offers several key advantages. First, it solely relies on the standard periodic MoM, so that it can be easily combined with existing in-house MoM software packages. Given the popularity of the MoM for the analysis of metasurfaces~\cite{MoM_metasurf_1, MoM_metasurf_2, MoM_metasurf_3, MoM_metasurf_4, MoM_metasurf_5}, having the possibility to study both the patches and the metasurface using the same code will greatly simplify the procedure. Second, the proposed technique does not entail any simplifying hypothesis. It rigorously models the propagation of the field along the periodic structure, handling both surface and space waves \cite{space_waves_1, space_waves_2, space_waves_3}. Third, thanks to the fast interpolation technique described, the computational cost of the proposed technique compares well with commercial softwares.

The remainder of this paper is organized as follows. In Section \ref{sec:fastMoM}, a fast MoM procedure to compute the periodic impedance matrices is presented (Section \ref{sec:perimdmat}). Then, the interpolation procedure and MBF-based compression are described (Section \ref{sec:interp}). In Section \ref{sec:TrackAlgo}, we present the fast tracking algorithm. In section \ref{Sec:results}, the proposed method is successfully applied to metasurface-type lossless periodic structures. The treatement of lossy structures or leaky modes requiring additional care, examples featuring conduction or radiation losses are presented in Section \ref{sec:lossy_examples}. It demonstrates the potential of the proposed technique to deal with damped modes and radiation.

\section{Fast MoM solution}
\label{sec:fastMoM}
To map the determinant of the impedance matrix on a regular grid, the periodic MoM has been accelerated using several techniques. First, the computation of the periodic impedance matrix at a few reference locations has been accelerated through the tabulation of the periodic Green's function in layered media. Then, using the reference impedance matrices, a specialized interpolation model is devised. Last, the size of the matrix is reduced using MBFs. These different steps are described in this section.

\subsection{Computation of the periodic impedance matrix}
\label{sec:perimdmat}
The periodic MoM impedance matrix is computed through convolution of a tabulated periodic Green's function (PGF) with the basis and testing functions. The accuracy is improved by using a singularity extraction technique similar, but not identical, to \cite{FLO15}. 

The PGF of layered media is generally computed in the spectral domain, where its value is known analytically \cite{POZ84}. However, the convergence of the spectral series may be slow. Moreover, its tabulation in the spatial domain is challenging due to the Green's function spatial singularity. 
To accelerate the convergence of the spectral series, the PGF is divided into two parts: the PGF in a homogeneous equivalent medium and a corrective term \cite{Jha_2014}. The permittivity of the equivalent homogeneous medium corresponds to the average between that of the dielectric medium and the upper medium (usually air), such that the spectral series of the corrective term converges rapidly. The homogeneous medium PGF is evaluated using a ``line-by-line" approach, described in \cite{GUE09}. The fields generated by each line of infinitesimal dipoles are expressed using a cylindrical-wave expansion. For each term of the expansion, the contribution of different lines are added up in space domain, with an acceleration based on the Levin-T formula \cite{Levin_T}. To limit the number of required cylindrical harmonics, the contribution of the line closest to the observation point is estimated in the spatial domain, with the use of the same accelerator. The contribution of the closest dipole is excluded to ease the tabulation of the PGF. 

Once the PGF has been tabulated, it is numerically convolved with the basis and testing function in the space domain \cite{CRA04}. The contribution of the closest dipole, which was excluded from the PGF, is evaluated in the spatial domain using standard (non-periodic) MoM, benefiting from well established singularity extraction procedures \cite{WIL84}. Since only the singularity associated to the closest source has been extracted, when basis and testing functions are almost one unit cell apart, the singularity stemming from a neighboring source may appear; that singularity is avoided by shifting the PGF by one period while introducing a corrective phase shift.

Using this method, the interaction between a periodic array of basis functions and a given testing function can be computed for any combination of phase shifts and frequency.  However, the resulting matrix is not a smooth function. This is why a new scheme is described in the next section, in which the singular terms are isolated and computed analytically, while the remainder is interpolated versus phase shifts.

\subsection{Interpolation of the periodic impedance matrix}
\label{sec:interp}
The interpolation procedure described here is an extension of \cite{D2, D3} by some of the authors to include interpolation versus frequency. It is based on the observation that sharp variations of the impedance matrix can be traced back to the sharp variation of a small number of Floquet modes in the infinite spectral series. Thus, the contribution of the troublesome Floquet modes is computed explicitly and the remainder is interpolated using a reduced number of reference points. The method shows some similarities with \cite{D6}, which was proposed by Z. Wang and S. Hum. However, there are some major differences. First, the proposed method also deals with the interpolation vs. phase shifts. Second, it is based on a less restrictive set of hypotheses, reducing the number of sampling points required for the interpolation. Last, the interpolation model is directly built on a reduced set of impedance matrices with no hypothesis on how the latter were computed. In this way, one does not need to explicitly evaluate the slowly converging infinite spectral series. Efficient hybrid spatial-spectral formulations can be used to compute the MoM periodic impedance matrices, significantly improving the computation time. In this work, we used the method described in Section \ref{sec:perimdmat}.

\begin{figure}
    \centering
    \includegraphics[width = 6cm]{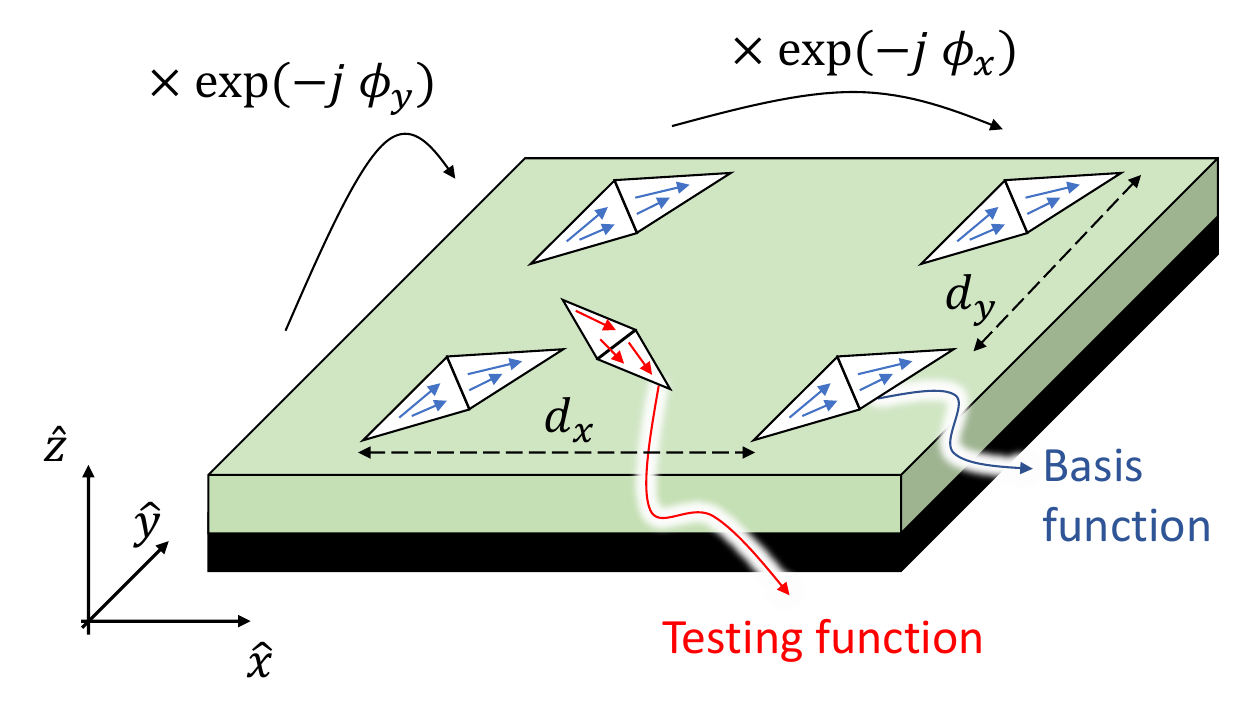}
    \caption{Geometry considered for the interpolation procedure}
    \label{fig:geom}
\end{figure}

Let us consider the interaction between a quasi-periodic array of basis functions (BFs) and a testing function (TF) above a substrate, as illustrated in Fig. \ref{fig:geom}. The array is quasi-periodic in directions $\dir{x}$ and $\dir{y}$ parallel to the substrate, with periods $d_x$ and $d_y$. For simplicity, we consider orthogonal directions. However, generalization to non-orthogonal directions is straightforward. The BFs are identical within phase shifts $\phi_x$ and $\phi_y$ between consecutive unit cells in the $\dir{x}$ and $\dir{y}$ directions, respectively. The direction normal to the substrate is denoted as $\dir{z}$, with positive $z$ above the substrate.  For simplicity, we consider that there is no overlap between the BFs and TF in the direction $\dir{z}$, i.e. $\exists z_0$ such that $\forall z \in \text{BF and } z' \in \text{TF}, (z-z_0)\times(z'-z_0) \leq 0$. This condition is, for example, automatically satisfied for flat structures printed on a substrate.

Using the Floquet theorem, the impedance matrix entry corresponding to the interaction between the BFs and the TF can be expressed as a sum of Floquet modes, i.e. plane waves whose transverse wavevectors $\vect{k}_{t,pq} = (k_{x,p}, k_{y,q}, 0)$ satisfy the quasi-periodic condition:
\newcommand{\kt}{\vect{k}_t}
\begin{align}
\label{eq:09}
k_{x,p} &= \dfrac{\phi_x + p 2\pi}{d_x} \\
k_{y,q} &= \dfrac{\phi_y + q 2\pi}{d_y}
\end{align}

To simplify the calculations, for a given plane wave, we consider the $\dir{e}$, $\dir{m}$ and $\dir{k}$ directions defined as:
\begin{align}
\dir{e} &= \dfrac{1}{k_t} \big( -k_y, k_x, 0 \big) \label{eq:07}\\
\dir{m}^\pm &= - \dfrac{1}{k_0 k_t} \big( \pm \kzup k_x, \pm \kzup k_y, - k_t^2 \big) \label{eq:08}\\
\dir{k}^\pm &= \dfrac{1}{k_0} \big(k_x, k_y, \pm \kzup)
\end{align}
with $k_0$, the free-space wavenumber, $k_t = \sqrt{\vect{k}_t \cdot \vect{k}_t}$ the transverse wavenumber, $\kzup = \sqrt{k_0^2-k_t^2}$ the vertical (upward) wavenumber. The $+$ sign corresponds to the plane waves directed upward (i.e. toward the $\dir{z}$ direction) and the $-$ sign corresponds to plane waves directed downward. Note that, in this paper, the square root function is defined such that its phase lies in $]-\pi, 0]$, such that the imaginary part of $\kzup$ is $\leq 0$. Using this special set of directions, the periodic impedance matrix describing the electric field generated on the TF by electric current on the periodic BFs can be expressed as
\newcommand{\Gammate}{\Gamma^\text{TE}}
\newcommand{\Gammatm}{\Gamma^\text{TM}}
\begin{equation}
\label{eq:01}
\begin{split}
Z^{J \rightarrow E} &= \dfrac{\eta_0 k_0}{d_x d_y} \sum_{p=-\infty}^\infty \sum_{q = -\infty}^\infty \dfrac{1}{2 \kzuppq} 
\\ &
\times \Big( 
\tilde{f}_{B,e}(\vect{k}_{pq}^\pm) \tilde{f}_{T,e}(-\vect{k}_{pq}^\pm) 
\\ & ~~~ 
+ \tilde{f}_{B,e}(\vect{k}_{pq}^-) \tilde{f}_{T,e}(-\vect{k}_{pq}^+) \Gammate(\vect{k}_{t,pq})
\\ & ~~~
+ \tilde{f}_{B,m^\pm}(\vect{k}_{pq}^\pm) \tilde{f}_{T,m^\pm}(-\vect{k}_{pq}^\pm) 
\\ & ~~~
+ \tilde{f}_{B,m^-}(\vect{k}_{pq}^-) \tilde{f}_{T,m^+}(-\vect{k}_{pq}^+) \Gammatm(\vect{k}_{t,pq}) \Big)
\end{split}
\end{equation}
where $\eta_0$ is the free-space impedance, $\kzuppq$ is the $\kzup$ wavenumber associated to Floquet mode $(p,q)$, $\Gammate$ and $\Gammatm$ correspond to the reflection coefficients of TE and TM waves on the substrate and the Fourier transform of the BF and TF are defined as 
\newcommand{\rpos}{\vect{r}}
\newcommand{\ktpq}{\vect{k}_{t,pq}}
\begin{equation}
\label{eq:03}
\tilde{f}_{B/T,i}(\vect{k}) = \dir{i} \cdot \iint \vect{f}_{B/T}(\rpos) \exp\big(j \vect{k} \cdot \rpos\big) d\rpos
\end{equation}
with $\vect{f}_B(\rpos)$ and $\vect{f}_T(\rpos)$ as the value of the BF and TF at position $\rpos$. Unit vector $\dir{i}$ can be either $\dir{e}$ or $\dir{m}$ (\eqref{eq:07} or \eqref{eq:08}).
Note that the first and third terms in \eqref{eq:01} correspond to the contribution of TE and TM waves to the free-space interaction between the BFs and TF, and the sign should be chosen to correspond to the geometry (BFs above or below the TF). For BFs and TFs on the same plane, any sign can be chosen.

Similarly, the magnetic field generated on the TF by electric currents on the BF reads
\begin{equation}
\label{eq:02}
\begin{split}
Z^{J \rightarrow H} &= \dfrac{k_0}{d_x d_y} \sum_{p=-\infty}^\infty \sum_{q = -\infty}^\infty \dfrac{1}{2 \kzuppq} 
\\ &
\times \Big( 
 \tilde{f}_{B,e}(\vect{k}_{pq}^\pm) \tilde{f}_{T,m^\pm}(-\vect{k}_{pq}^\pm) 
\\ & ~~~ 
+ \tilde{f}_{B,e}(\vect{k}_{pq}^-) \tilde{f}_{T,m^+}(-\vect{k}_{pq}^+) \Gammate(\vect{k}_{t,pq})
\\ & ~~~
- \tilde{f}_{B,m^\pm}(\vect{k}_{pq}^\pm) \tilde{f}_{T,e}(-\vect{k}_{pq}^\pm) 
\\ & ~~~
- \tilde{f}_{B,m^-}(\vect{k}_{pq}^-) \tilde{f}_{T,e}(-\vect{k}_{pq}^+) \Gammatm(\vect{k}_{t,pq}) \Big)
\end{split}
\end{equation}

To efficiently interpolate the impedance matrix, a good understanding of the evolution of $Z^{J \rightarrow E}$ and $Z^{J\rightarrow H}$ with phase shifts and frequency is required. This paper concentrates on the analysis of $Z^{J \rightarrow E}$. However, $Z^{J \rightarrow H}$ can be treated similarly. 

A detailed analysis of $Z^{J \rightarrow E}$ (cf. Appendix \ref{app:evol_Z}) shows that the contributions of a limited set of low-order Floquet modes can exhibit sharp variations and is thus difficult to interpolate. However, the contribution of high-order Floquet modes evolves smoothly with frequency and phase shifts. Moreover, for small BF and TF, evolution with frequency is roughly proportional to $1/k_0$ while evolution with phase-shift corresponds to a linear phase shift proportional to the relative positions of the BF and the TF. 

Exploiting these observations, an efficient interpolation method can be devised. First, as classically done (see e.g. \cite{Reviewers_1, Reviewers_2, D6}), the total impedance matrix is split into two parts: one corresponding to visible and slightly evanescent Floquet modes (i.e. of orders below chosen thresholds $N_x, N_y$: $-N_x \leq p \leq N_x, -N_y \leq q \leq N_y$), denoted as $\Zje_v$, and one corresponding all the other Floquet modes, denoted as $\Zje_e$:
\begin{equation}
\label{eq:05}
\Zje(k_0, \phi_x, \phi_y) = \Zje_v(k_0, \phi_x, \phi_y) + \Zje_e(k_0, \phi_x, \phi_y)
\end{equation}
The values of $N_x$ and $N_y$ depend on the dimensions of the unit cell, and are typically of the order of $0~... ~2$ for a metasurface. Provided that $N_x$ and $N_y$ are small enough, the computation of the the first term is very rapid: the reflection coefficient is independent from the pair of BF and TF considered and is thus only computed once, and the Fourier transforms of the BF and TF can be evaluated separately and combined afterwards. Note that the latter is true only because we considered BF and TF with no overlap in the $\dir{z}$ direction. If there is an overlap, the formulation of \eqref{eq:01} should be slightly modified, so that the factors related to the BF and TF are no longer separable for the free-space interaction. 

\begin{figure*}
\centering
\includegraphics[width = 14cm]{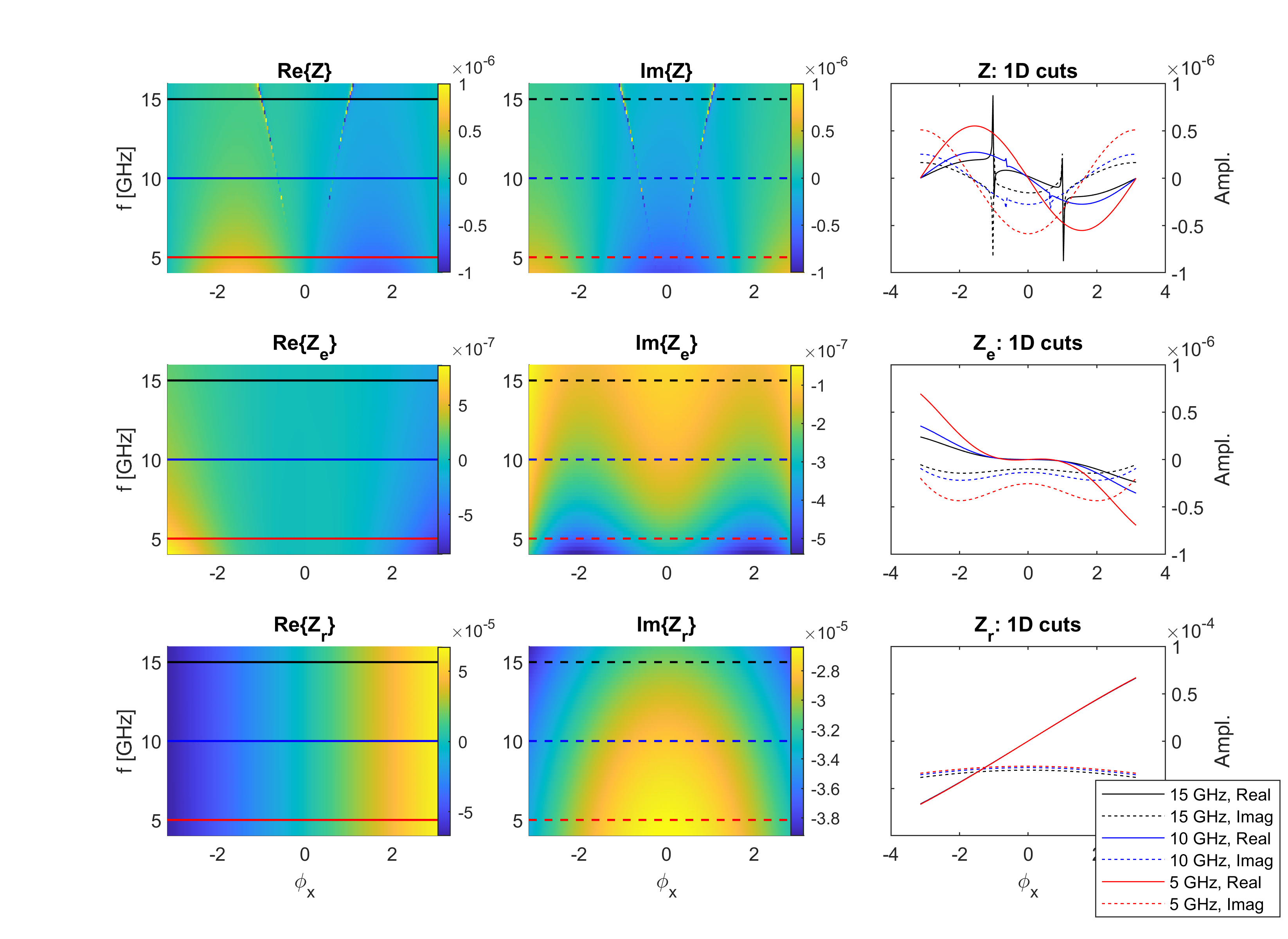}
\caption{Evolution of the $\Zje$ (top), $\Zje_e$ (middle) and $\Zje_r$ (bottom) terms with frequency and phase shift for a typical pair of Rao-Wilton-Glisson basis and testing functions. Left: real part of the function. Center: imaginary part of the function. Right: 1D cuts at 5 GHz, 10 GHz and 15 GHz. The continuous line corresponds to the real part and the dashed line to the imaginary part. The progressive smoothing of the function with each consecutive step is clearly visible.}
\label{fig:evol_Z}
\end{figure*}

Concerning the second term of \eqref{eq:05}, evaluating it using the infinite series of Floquet modes can be time consuming. Moreover, the absence of low-order Floquet modes can impede the use of specialized acceleration techniques, such as the one described in Section \ref{sec:perimdmat}. However, it can be easily interpolated after few minor modifications, namely multiplication by a phase factor and by the frequency. Thus, the raw periodic impedance matrix $\Zje$ is evaluated on a few phase shifts and frequencies. Then, for each sampling point, $\Zje_v$ is evaluated and subtracted from the total impedance matrix. Last, a phase term proportional to the average lateral distance $\Delta x$ and $\Delta y$ between the BF and TF is removed, the result is multiplied by $k_0$ and the remainder, denoted as $\Zje_r$, is interpolated:
\begin{equation}
\label{eq:04}
\begin{split}
\Zje_r = k_0 \exp\bigg(j  \Big(\dfrac{\phi_x \Delta x}{d_x} &+ \dfrac{\phi_y \Delta y}{d_y}\Big)\bigg) \\
& \times \big(\Zje - \Zje_v\big)
\end{split}
\end{equation}
Interpolation is carried out using a multidimensional polynomial. Excellent results are obtained due to the very smooth function obtained after removing the contribution of the main Floquet modes, of the phase factor and of the inverse frequency. As an illustration, the evolution of the $\Zje$, $\Zje_e$ and $\Zje_r$ terms with frequency and phase shift is illustrated in Fig. \ref{fig:evol_Z} for a typical pair of BF and TF separated by half a unit cell.

Once the polynomial interpolation model for $\Zje_r$ has been built, the interpolation procedure can be summarized as follows. First, the $\Zje_r$ term is interpolated. Then, the $\Zje_v$ term is computed as a small series of Floquet modes. The impedance matrix entry is then obtained using
\begin{equation}
\label{eq:06}
    \Zje = \Zje_v + \dfrac{\Zje_r}{k_0} \exp\bigg(-j  \Big(\dfrac{\phi_x \Delta x}{d_x} + \dfrac{\phi_y \Delta y}{d_y}\Big)\bigg)
\end{equation}
Using this procedure, the periodic impedance matrix (and its determinant) can be rapidly evaluated over a fine grid of phase shifts and frequencies. To further accelerate the technique, three computational bottlenecks have been identified and alleviated/solved. First, to build a 3D polynomial interpolation model, a large number of reference matrices need to be computed. Second, when interpolating the impedance matrix using \eqref{eq:06}, most of the time is devoted to the evaluation of the Fourier transform of the BF and TF, even when analytical formulas are available (for example in the case of RWG~\cite{RAO82} or rooftop BF and TF). Last, for large meshes, the interpolation procedure asymptotically scales as the square of the number of unknowns, and the evaluation of the determinant as its cubic power, rapdily increasing the computation time with the mesh density.

To mitigate the first bottleneck, we exploited the $2\pi$ periodicity of the periodic impedance matrix versus $\phi_x$ and $\phi_y$. Since $\Zje$ is periodic, once $\Zje(\phi_x, \phi_y)$ has been computed, $\Zje(\phi_x + 2m \pi, \phi_y + 2n \pi)$ does not provide any additional information and thus cannot be used as such to improve the interpolation model. However, the choice of the Floquet modes that are removed depends on the phase shifts considered since $k_{x,p}(\phi_x) = k_{x,(p-1)}(\phi_x + 2\pi) \neq k_{x,p}(\phi_x + 2\pi)$ (see Equation \ref{eq:09}). Thus, $\Zje_r$ is not periodic: its values at positions $\Zje(\phi_x, \phi_y)$ and $\Zje(\phi_x + 2m \pi, \phi_y + 2n \pi)$ (with $m$ or $n$ different from zero) convey complementary information and can thus be used to improve the interpolation model. Thus, the same periodic matrix can be used to extract the value of $\Zje_r$ at several different phase shifts separated by a distance $2\pi$. Even if, eventually, the interpolation model of $\Zje_r$ only needs to cover the first Brillouin zone, it can be improved using sampling points located outside of this zone provided that $\Zje_r$ is smooth over the extended zone. Thus, there is a trade-off that can be optimized: the zone over which $\Zje_r$ is smooth can be extended by increasing $N_x$ and $N_y$, reducing the number of periodic impedance matrices required to build the model. However, the cost of evaluating $\Zje_v$ increases with $N_x$ and $N_y$, which impacts all the subsequent interpolations. In practice, we observed that slightly increasing $N_x$ and $N_y$ (from $N_x, N_y = 0$ to $N_x, N_y = 1$ in the examples considered) and using sampling points over the zone $[-2\pi, 2\pi]$ was beneficial to the accuracy and total computation time. For a given frequency, only 4 periodic impedance matrix $\Zje$ were required to build a 2-D 4th order polynomial model!

To remove the second bottleneck and accelerate the evaluation of the Fourier transform of the BFs and TFs, we used a polynomial interpolation technique. First, an average phase term related to the mean position of the BF or TF is removed. Then the remainder is fitted using a polynomial model. For BFs and TFs that are not horizontal, special care is required to handle the angular point related to the visibility limit, and thus the angular behaviour of $\kzup$. However, for co-planar horizontal BF and TF, the $\kzup$ factor has no impact on the interpolant, simplifying the interpolation procedure. In the geometries studied, an excellent accuracy has been obtained with relatively low polynomial orders (around order 5), further accelerating the computations.

Last, to reduce the number of unknowns, a Model Order Reduction (MOR) \cite{D19, D1} technique based on MBFs \cite{D7bis} has been used. First, using the interpolation model, we compute the equivalent currents distribution generated on the periodic structure by incident plane waves with various transverse wave vectors and at various frequencies. A reduced basis that can reproduce all the current distributions within a prescribed accuracy is built. Each vector of this new basis (i.e. MBF) corresponds to a linear combination of BFs. Projecting the full impedance matrices into the new basis, one obtains a reduced impedance matrix that can serve as an accurate approximation of the original one. More information about the procedure can be found in Appendix \ref{app:mbfs}. 

Once the set of MBFs has been obtained, the interpolation procedure described previously is applied to the reduced matrices. On one hand, the reduced number of unknowns leads to smaller matrices, i.e. less entries that need to be interpolated and a faster evaluation of the determinant. On the other hand, MBFs have a larger spatial extent than BFs, so that their Fourier transforms are less smooth. Higher-order interpolation models are thus required to interpolate the Fourier transform of the MBFs and the $\Zje_r$ terms. In the examples studied, the use of MBFs accelerated the mapping of the determinant by a factor ranging from 3.5 to 8. This factor accounts for the time devoted to the computation of the MBFs.

\subsection{Summary of the algorithm}
The algorithm used to interpolate the reduced impedance matrix is summarized below. The algorithm has been split into two parts. First, the preparation (P) corresponds to the computation of the MBFs and the establishment of the interpolation model. It only needs to be carried out once for a given geometry. Then, the interpolation (I) itself, must be repeated for each new combination of frequency and phase shifts at which the reduced impedance matrix is required. 

\underline{\textbf{Preparation}}
\begin{itemize}
\item[P1] Build an interpolation model for the Fourier transform of the BFs and TFs;
\item[P2] Sample the interpolant of the periodic impedance matrix ($\Zje_r$) at several phase shift and frequencies, i.e:
	\begin{itemize}
	\item[P2.1] Compute the periodic impedance matrix (or use a previously computed matrix);
	\item[P2.2] Remove the contributions of the low-order Floquet modes;
	\item[P2.3] Remove the phase factor and multiply by the frequency;
	\end{itemize}
\item[P3] Build a polynomial model that can approximate the interpolant;
\item[P4] Build the MBFs;
\item[P5] Sample the interpolant ($\Zje_r$) of the reduced impedance matrix for several phase shifts and frequencies, i.e:
	\begin{itemize}
	\item[P5.1] Interpolate the $\Zje_r$ term of the full-size impedance matrix;
	\item[P5.2] Add the phase factor;
	\item[P5.3] Project the matrix into the MBFs;
	\item[P5.4] Remove the phase factor associated with the MBFs;
	\end{itemize}
\item[P6] Build a polynomial model that can approximate the new interpolant;
\item[P7] Build an interpolation model for the Fourier transform of the MBFs.
\end{itemize}

\underline{\textbf{Interpolation}}
\begin{itemize} 
\item[I1] Interpolate the $\Zje_r$ term of the MBF-compressed impedance matrix;
\item[I2] Multiply by the phase factor and divide by the frequency;
\item[I3] Add the contributions of the low-order Floquet modes;
\item[I4] Compute the logarithm of the determinant of the reduced impedance matrix.
\end{itemize}
Note that, when MBFs are defined over the entire metallization, the average distance between two different MBFs is zero, so that the phase factors mentioned in steps P5.4 and I2 simply correspond to a factor of 1.

\section{Tracking Algorithm}
\label{sec:TrackAlgo}
\subsection{Choice of the Metric}
\label{sec:TrackAlgoMetric}
To track the dispersion curves, we chose to map the determinant of the periodic impedance matrix. Tracking the determinant has several advantages and disadvantages over other metrics, such as the conditioning number or the so-called \textit{characteristic term} \cite{D4bis}. On one hand, using the determinant, it is possible to distinguish between zeros and poles of the matrix (i.e. diverging or vanishing eigenvalues) or to use complex-analysis based techniques since the value of the determinant is an analytic function of the entries of the matrix. On the other hand, if the number of unknowns is large, the computation of the determinant can be costly and prone to overflow or underflow error using standard arithmetic precision. 

For the geometries studied, the computational cost was found to be acceptable thanks to the use of MBFs. To avoid overflow or underflow error, we chose to compute the logarithm of the determinant instead of the determinant itself. The amplitude and phase of the determinant are directly related to the real and imaginary parts of its logarithm, respectively. To compute the logarithm of the determinant without explicitly computing the determinant, the following identity has been used:
\begin{equation}
\det\{A\} = k^N \det\{A/k\}
\end{equation}
with $A$ a $N$x$N$ matrix (here, $N$ is the number of MBFs) and $k$ a scalar. The logarithm of the determinant then becomes
\begin{equation}
D_r + j D_i = \log\big(\det\{A\}\big) = N \log(k) + \log\big(\det\{A/k\}\big) 
\label{eq:dd}
\end{equation}

The actual value of the logarithm was found iteratively. First, a factor $k$ is used to ``rescale" the matrix. The value of $k$ can be obtained from \textit{a-priori} information, such as the value of the determinant for frequencies or phase shifts that are close. If the determinant of the scaled impedance matrix is \texttt{inf} (resp. \texttt{0}), the $k$ factor is rescaled by a factor $10^{-280/N}$ (resp. $10^{280/N}$), with $10^{280}$ that is slightly smaller than the maximum number that can be represented using standard \texttt{double} precision, and $N$ the size of the impedance matrix. Using this method, the value of the determinant is generally obtained after few iterations when no a-priori information is available, and at the first iteration otherwise.

\subsection{Computation of Dispersion Surfaces}
\label{sec:ModelDisperSion}
\begin{figure}
    \centering
    \includegraphics[width = 8.8cm]{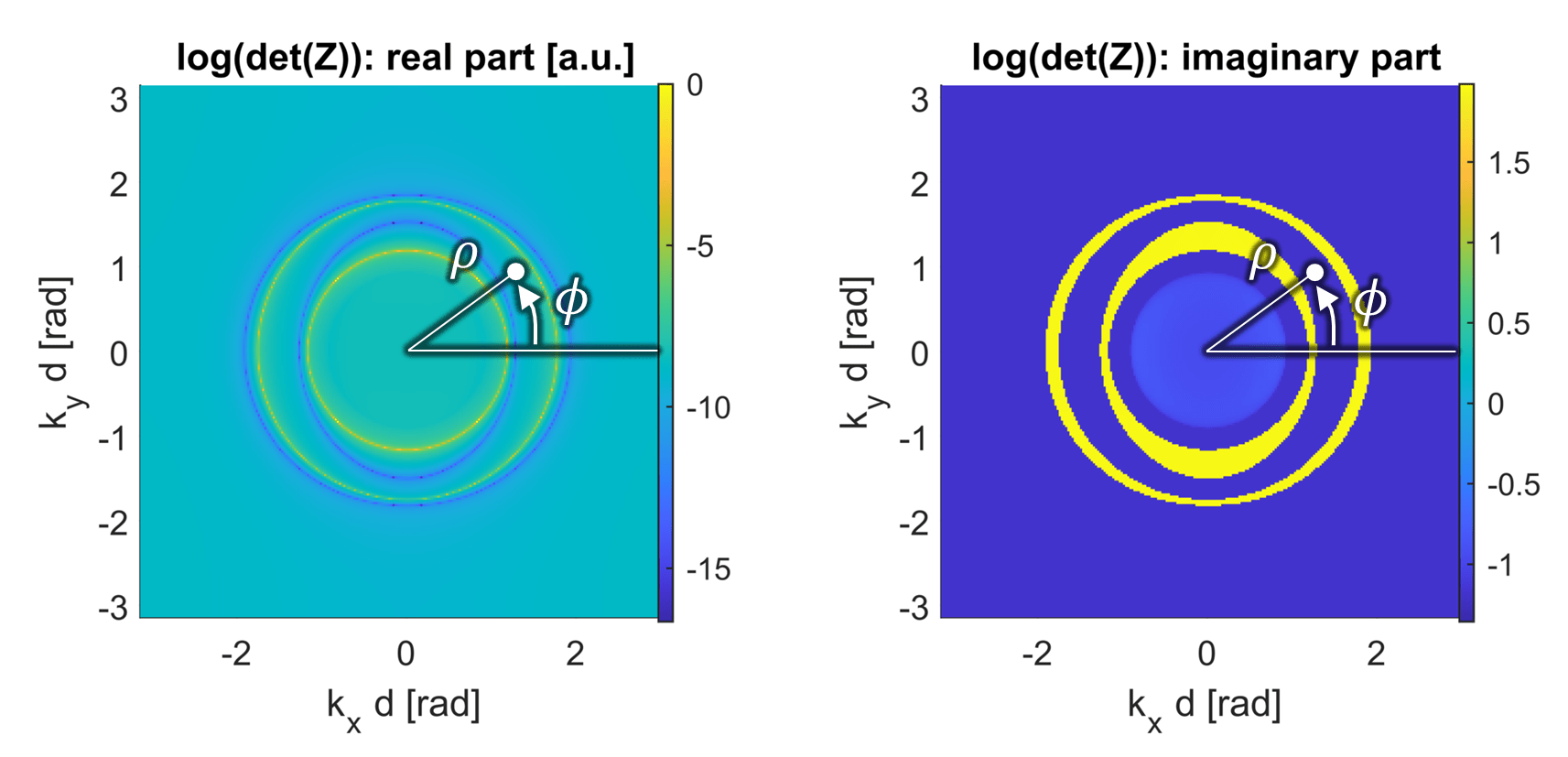}
    \caption{Map of the real (left) and imaginary (right) parts of the logarithm of the determinant for a typical geometry. Poles and zeros of the determinant are characterized by an increase (yellow) or decrease (blue) of the real part of its logarithm, respectively, and a sharp change of its imaginary part.}
    \label{fig:PixelRadRho}
\end{figure}

For lossless surface waves, the dispersion surfaces in the $(f, \phi_x, \phi_y)$ space are computed as stacks of iso-frequency contours in successive $(\phi_x, \phi_y)$ planes. In each plane, the logarithm of the determinant of the reduced impedance matrix is mapped. Iso-frequency curves are associated with a vanishing determinant, i.e. a real part of its logarithm going to $-\infty$. A typical map of the real and imaginary parts of its logarithm for a lossless open periodic structure is provided in Fig. \ref{fig:PixelRadRho}. Different features can be observed. First, inside the visible region, the phase of the determinant can vary smoothly, due to the presence of radiation loss. Since the frequency and phase shifts are real, no sharp variation is expected, poles and zeros being located at complex coordinates (i.e. complex frequencies and/or phase shifts). Outside the visible zone, there is no radiation loss, so that sharp variations of the amplitude and phase associated with zeros and poles of the determinant are visible. Poles are associated with surface waves of the substrate (i.e. poles of the PGF), while zeros are associated with iso-frequency contours. In addition to poles and zeros, features associated to Wood's anomalies (i.e. branch points in the PGF due to grazing incidence of the main or secondary lobes) may appear.

The determination of iso-frequency contours developed here assumes a closed curve in the ($\phi_{x}, \phi_{y}$) plane. Using polar coordinates ($\rho, \phi$), each contour is parameterized using a harmonic model versus angular coordinate $\phi$:
\begin{equation}
\rho_{i}(\phi) = c_i + \sum_{n} \big({d}_{i,n} \cos(n\phi)  + e_{i,n} \sin(n\phi)\big), 
\label{isocurvemodel}
\end{equation}
with $i$ corresponding to the index of the curve. 

To find the coefficients of the harmonic expansion, a weight is attributed to each pixel of the map, with larger weights attributed to pixels that are expected to be close to the curve. 

The weight depends on several indicators. The first one obviously depends on the log-magnitude of the determinant, corresponding to $D_r(\phi_x,\phi_y)$ equ.~\eqref{eq:dd}. It is taken as $D_r(0,0) - D_r(\phi_x,\phi_y)$. The next indicator corresponds to the local variation of the determinant versus phase shifts $\phi_x$ and $\phi_y$. More precisely, it is computed as $|\partial D_r / \partial \psi_x|+|\partial D_r / \partial \psi_y|$. If a previous estimate of the model, $\rho_p(\phi)$, is available (see below), the proximity to the model also appears as a useful indicator, e.g. $\exp(-|\rho(\phi)-\rho_p(\phi)|^2/(2 \sigma^2))$ where $\sigma$ is smaller when the previous estimation is deemed more reliable. The first two indicators are multiplied with each other
and the result should be larger than a given threshold $\mu$.
The third indicator is used to exclude pixels that are too far from a previous estimate if available.
Using these weights, the determination of coefficients $c_i$, $d_{i,n}$ and $e_{i,n}$ becomes a weighted least-squares problem, the goal being to minimize the RMS radial distance between each pixel and the curve. Mathematically, it amounts to solving the over-constrained system of equations:
\begin{equation}
w\cdot M\,x = w \cdot \rho
\end{equation}
where each line corresponds to a given pixel of the map. $w$ is a column vector containing the weights associated to each pixel, $x$ is a column vector containing all the model coefficients $(c_i, d_{i,n}, e_{i,n})$, $\rho$ is a column vector containing the radial coordinate of the pixels and $M$ is a matrix whose columns correspond to vectors $1$, $\cos(n\phi)$ and $\sin(n\phi)$ for all $n$. The dot multiplication ($\cdot$) here means column-wise element-by-element multiplication. Pixels with a low weight, typically below threshold $\mu=0.16$, can be excluded from the system of equations to reduce the computation time. It can be noted that some other parameters may have to be tuned based on the type and the complexity of patches under consideration. For instance, in the case of rectangular patches (Fig. \ref{fig:rectangles1}) the harmonic order needed is $2$, while for the case of a more complex structure, such as coffee bean patch (Fig. \ref{fig:coffee_bean}), for which several contours appear, that order needs to be increased to $6$. In terms of the $\sigma$ parameter, 
related to proximity with previous estimates,
for both cases, it remains identical and is equal to $0.04\pi$ rad. 

In case two contours are observed, the outer one is first modelled. Larger weights are  provided to pixels of the outer contour (as compared to pixels of the inner contour) by providing larger weights for larger values of $\rho$. Then, an iterative refinement only accounting for pixels close to the previous estimate can be used to better catch the outer curve. Once the position of the outer curve has been estimated with a sufficient precision, the inner curve position can be determined by excluding the contribution of the pixels that are close to or outside the estimate of the outer curve. Iterating, the cross-influence between different curves can be rapidly eliminated.

\section{Numerical results: lossless case}
\label{Sec:results}
The interpolation method and the tracking algorithm have been tested on different arrays of printed patches: square arrays of rectangular patches with varying orientations and a square array of coffee-bean patches proposed in \cite{MTSforspace}, \cite{inverion_SW_antenna} for the design of leaky-wave metasurface antennas. 

The same numerical parameters have been used for two geometries to interpolate the periodic impedance matrix. To build the interpolation model of the impedance matrix itself, only three sampling points have been used versus frequency. For each frequency, only four periodic matrices associated to phase shifts $\phi_x, \phi_y \in \{0, \pi\}$ have been computed using the method described in Section \ref{sec:perimdmat}. From these 12 matrices, $\Zje_r$ has been estimated at 75 locations, for phase shifts ranging from $-2\pi$ to $2\pi$, by extracting the 9 dominant Floquet modes ($N_x = N_y = 1$). Using these points, the slowly varying term $\Zje_r$ (cf. Equation \eqref{eq:04}) was interpolated using a 3D polynomial of orders 4, 4 and 2 along the $\phi_x$, $\phi_y$ and $f$ dimensions, respectively. The polynomial interpolation of the Fourier transform of the BF and TF over the whole zone of interest was carried out using a 2D polynomial of orders 5 along both the $k_x$ and $k_y$ dimensions.

Once the interpolation model was built, the MBFs have been computed considering 3645 different plane wave excitations of various frequencies and wavevectors. The thresholds $\theta_1 = 10^{-3}$ and $\theta_2 = 10^{-15}$ have been chosen (cf. Appendix \ref{app:mbfs}), leading to 20 and 11 MBFs (depending on the orientation) in the case of the rectangular patches and 86 MBFs in the case of coffee bean patches. The interpolation of the reduced impedance matrix was done using polynomials of order 9, 9 and 4 along the $\phi_x$, $\phi_y$ and $f$ dimensions, respectively. Their Fourier transform was interpolated using polynomials of orders 11 along $k_x$ and $k_y$. Note that, using this set of parameters, the relative error introduced by the whole interpolation procedure on the entries of the reduced impedance matrix was estimated to be below $10^{-3}$.

\begin{figure}
\centering
\includegraphics[width = 8.8cm]{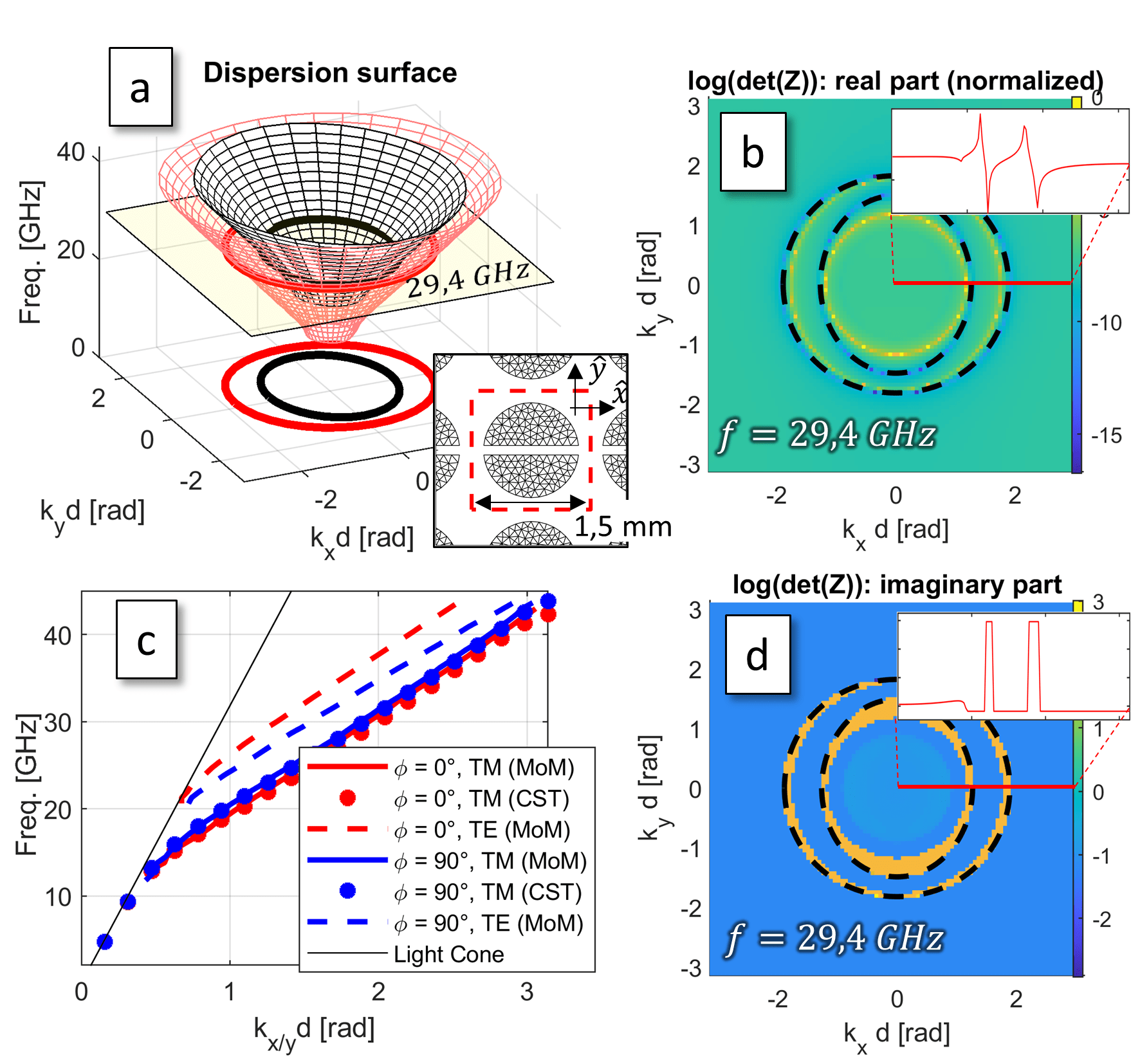}
\caption{(a): Dispersion surface of the coffee bean patches (see inset). (c): Dispersion curves of the (quasi-)TM and TE modes obtained using the proposed method and using CST. (b,d): Iso-frequency curves (black dashed lines) of the TE (inner curve) and TM (outer curve) modes at 29.4 GHz. They are superimposed to the map of (b) the real part and (d) the imaginary part of the logarithm of the determinant. Finely sampled transverse cuts along the direction $\phi=0$ are illustrated in the insets.}
\label{fig:coffee_bean}
\end{figure}

Last, the determinant has been mapped over a regular grid of $100 \times 100 \times 30$ points versus $\phi_x$, $\phi_y$ and $f$, respectively. The computation of the full dispersion surfaces in the real $\phi_x-\phi_y-f$ space took approximately 10 minutes for the rectangular patches (see Figs. \ref{fig:rectangles1} and \ref{fig:rectangles2}) and 15 minutes for the coffee-bean patches (see Fig. \ref{fig:coffee_bean}). For comparison, using the same computer, the computation of the reference results obtained using CST~\cite{CSTref} (see dots in Figs. \ref{fig:rectangles1}(b,d) and \ref{fig:coffee_bean}(c)) took approximately 15 and 5 minutes, respectively. It is important to note that CST results only correspond to 1D cuts of the full dispersion surfaces and that the CST computations were fully parallelized. 

The detailed timing of the whole procedure is provided in Table \ref{tab:timing}. The simulations were run on a \texttt{i7-10700} processor without any parallelization, except for the computation of the determinant using the \texttt{det} built-in Matlab function. It can be noticed that the time is dominated by the mapping of the determinant over a large number of points. This time could be reduced by adaptively choosing where the MoM impedance matrices need to be evaluated, as the zeros of the determinant are being tracked in the $f-\phi_x-\phi_y$ domain. Such refined techniques are out of the scope of the paper.

\begin{table}
\begin{center}
\begin{tabular}{ |L{4cm}||C{1.5cm}|C{1.5cm}|}
 \hline
 Mesh & 
    Rectangles (325~BF) &
    coffee beans (348~BF)\\
 \hline \hline
 \multicolumn{3}{|c|}{Preparation time [s]} \\
\hline
Computation of periodic impedance matrices (12) & 
    138 &
    155\\
\hline
Preparation of the interpolation models & 
    3 &
    3.1 \\
\hline
Computation of the MBFs & 
    12.5 &
    13.3\\
\hline
Preparation of the interpolation models for MBFs & 
    10 &
    21 \\
\hline
\textbf{Total} & 
    164 &
    192 \\
\hline\hline
 \multicolumn{3}{|c|}{Mapping of the determinant and tracking of the curves [s]} \\
\hline
Mapping & 
    263 &
    508\\
\hline
Tracking & 
      0.25 &
      16 \\
\hline
\end{tabular}
\\~
\caption{Time required to obtain the dispersion curves of the $\hat{x}$-directed rectangular patches and the coffee bean patches. \vspace{-0.3cm}}
\label{tab:timing}
\end{center}
\end{table}

\subsection{Rectangular patch}
\label{Rectangular patch}

\begin{figure}
\centering
\includegraphics[width = 8.8cm]{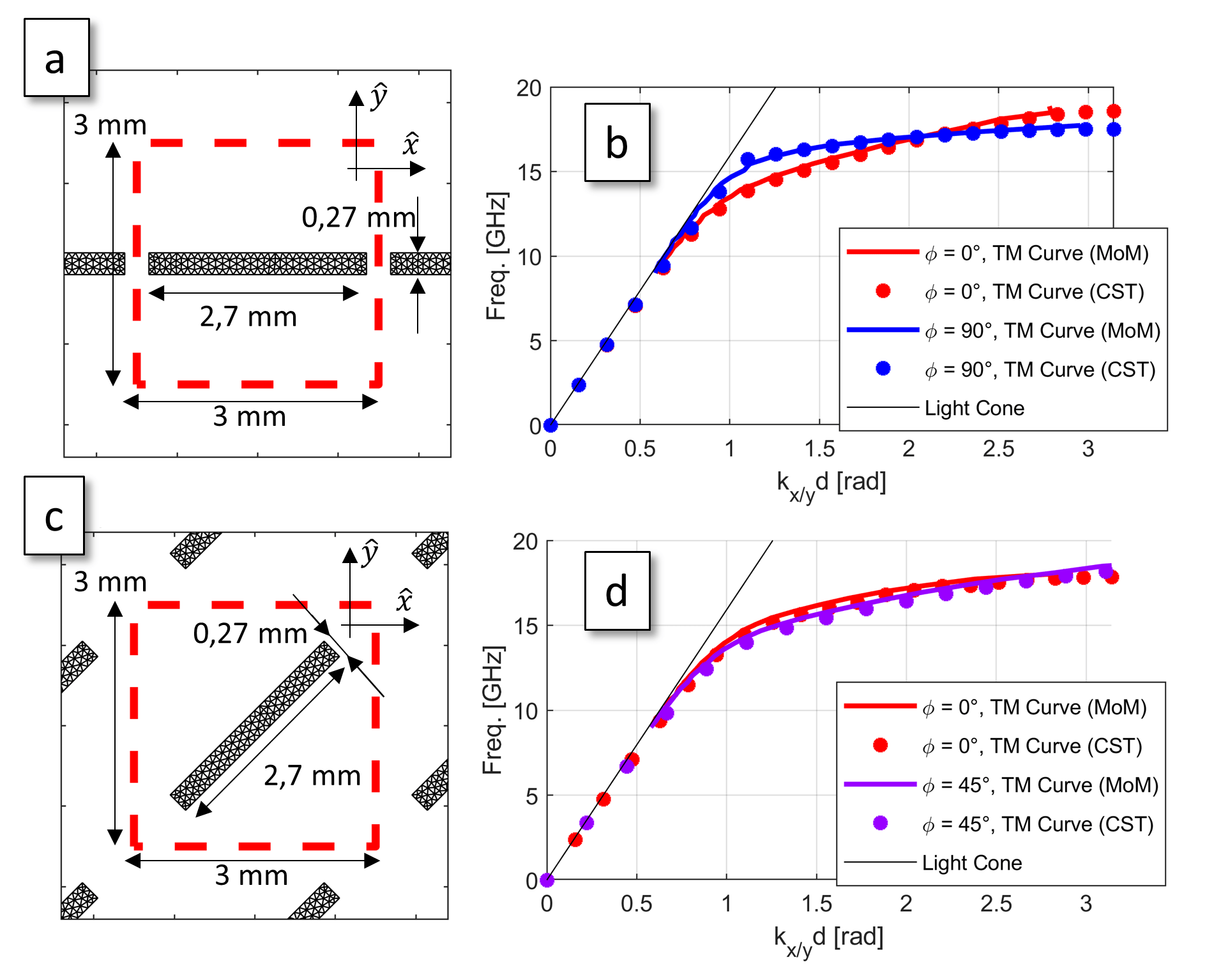}
\caption{(a) and (c): geometry of the two arrays of rectangular patches studied. (b) and (d): dispersion curves obtained using the method described in this work and the commercial software CST \cite{CSTref}.}
\label{fig:rectangles1}
\end{figure}

\begin{figure}
\centering
\includegraphics[width = 8.8cm]{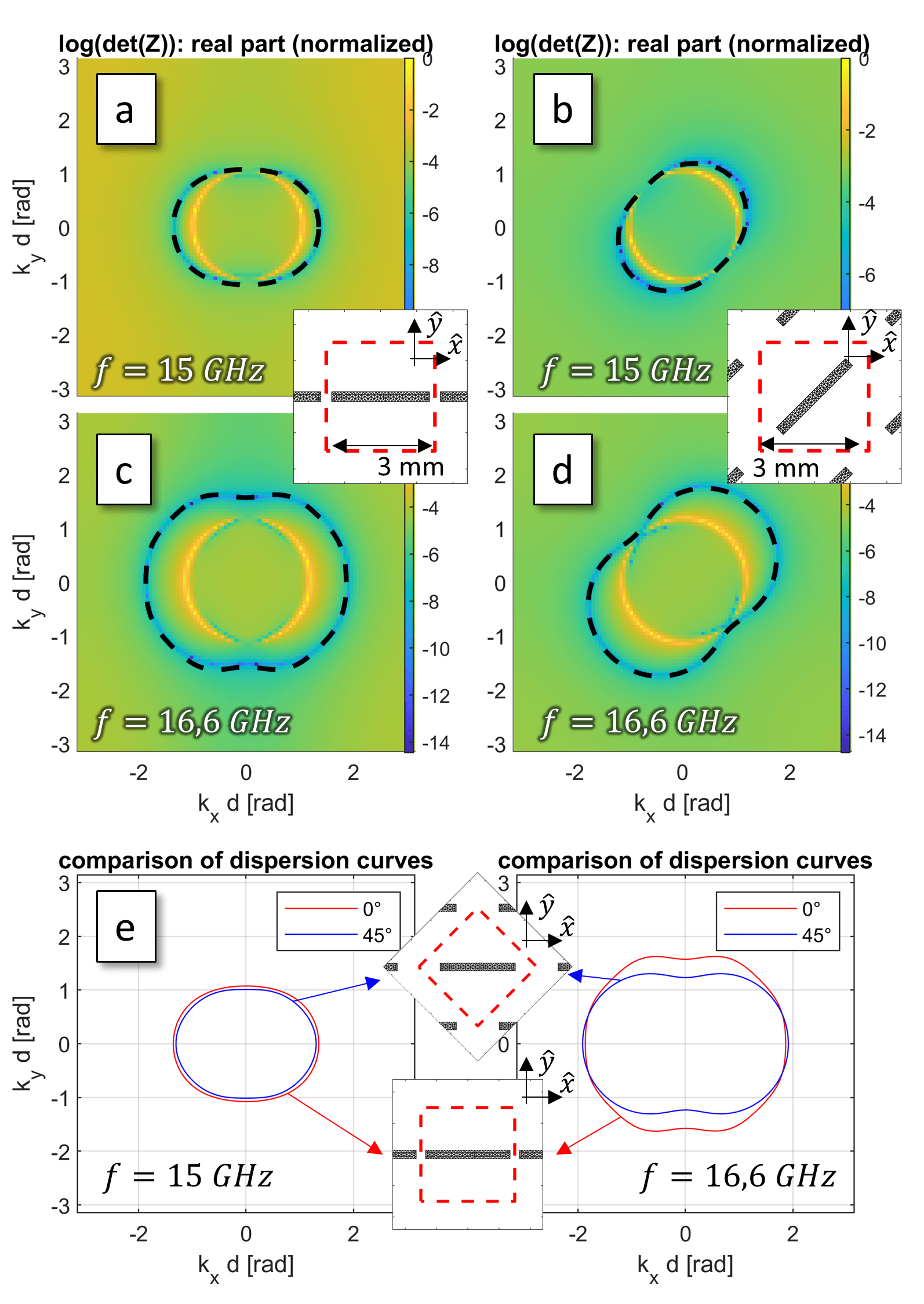}
\caption{Iso-frequency curves of the arrays of rectangular patches (a,c) aligned with the $\dir{x}$ direction and (b,d) rotated by $45^\circ$, 15 GHz and 16.6 GHz. The estimated dispersion curves (dashed black line) is superimposed to the map of the absolute value of the determinant, in logarithmic scale. (e) Comparison of the iso-frequency curves of both arrays. Note that the iso-frequecy cuve of the $45^\circ$-rotated patches has been rotated again to facilitate the comparison.}
\label{fig:rectangles2}
\end{figure}

The dispersion surfaces of periodic arrays of rectangular patches have been studied at frequencies ranging from 9 GHz to 20 GHz. The patches are deposited on a 1 mm-thick grounded dielectric substrate of relative permittivity 9.8. The patches have been discretized using 325 RWG basis functions. In one configuration, the patches are aligned with the direction of periodicity. In the other configuration, they are rotated by 45$^\circ$. Both geometries are illustrated in Fig. \ref{fig:rectangles1}(a,c). 

It is well known that, at low frequency, this structure supports the propagation of a dominantly TM surface wave (SW). The dispersion curves of both patches have been computed for a fixed azimuthal direction $\phi$ (see conventions in Fig. \ref{fig:PixelRadRho}) and validated using the eigenmode solver of the CST commercial software \cite{CSTref}. To use the eigenmode solver, the simulation zone has been truncated using a perfectly conducting plane located sufficiently far away from the periodic structure to prevent evanescent coupling between the structure and the plane. Since the surface mode does not radiate into free-space, accurate results could be obtained in this way. As shown in Fig. \ref{fig:rectangles1}(b,d), the dispersion curves obtained by both means match very well.

To look at the impact of the rotation of the patches, the iso-frequency curves of both arrays have been displayed in Fig. \ref{fig:rectangles2} at frequencies of 15 GHz and 16.6 GHz, along with the map of the logarithm of the amplitude of the determinant. As expected, the iso-frequency dispersion curve is less circular at higher frequencies, which means that the response of the structure is more anisotropic. To highlight the impact of the rotation of the patches, the iso-frequency curves of the two arrays are compared in Fig. \ref{fig:rectangles2}(e).
It is observed that at lower frequencies, the effect of the rotation of the patch can be accurately estimated by simply rotating the dispersion curve. However, moving to higher frequencies, this is no longer true because of the different relative positions of the patches.

It should be noted that, although our interpolation technique is valid at any frequency, the proposed eigenmode tracking method assumes a closed iso-frequency dispersion curve due to the harmonic expansion along azimuth used in Eq.~\eqref{isocurvemodel}. This assumption is only valid as long as the unit-cells are sufficiently subwavelength, limiting the maximum frequency. However, this condition is usually met for metasurfaces, in which the homogenization assumption imposes sub-wavelength unit cells. A different eigenmode tracking algorithm is required at higher frequencies, e.g. when studying bandgap materials.

\subsection{Coffee bean patch}
\label{Coffee bean patch}
A periodic array of coffee bean patches has been studied at frequencies between 10 GHz and 45 GHz. The patches correspond to circles of radius 600 $\mu$m that are split into two parts by a 120 $\mu$m slit. The patches have been discretized using 348 basis functions and are illustrated in the inset of Fig. \ref{fig:coffee_bean}(a). The patches are deposited on a 1.5 mm thick grounded substrate of relative permittivity 6, with a periodicity of 1.5 mm. This patch type has been proposed in \cite{MTSforspace}, \cite{inverion_SW_antenna} for the design of leaky-wave metasurface antennas. 

The dispersion surfaces of the patches are displayed in Fig. \ref{fig:coffee_bean}(a). It can be noticed that, at high frequency, two different surfaces are visible, corresponding to the quasi-TM (red) and quasi-TE (black) modes. The iso-frequency curves at 29.4 GHz have been extracted from the surfaces and are displayed in Figs. \ref{fig:coffee_bean}(b,d) (black dashed line). They are superimposed to the map of the real and imaginary parts of the logarithm of the determinant. As explained in Section \ref{sec:ModelDisperSion}, the dispersion curves correspond to zeros of the determinant and are thus associated with a phase jump of $\pi$ (Fig. \ref{fig:coffee_bean}(d)).

Cuts of the dispersion surface in the planes $\phi=0$ and $\phi=90^\circ$ are provided in Fig.~\ref{fig:coffee_bean}(c), as well as a comparison with CST results for the quasi-TM mode. A very good agreement is obtained.

\section{Numerical results: lossy case}
\label{sec:lossy_examples}
The surface modes can be damped due to the presence of radiation or conduction losses. The study of lossy surface modes is more delicate. The coordinates of the lossy modes in the $f-\phi_x-\phi_y$ space are complex, the imaginary part being related to the damping coefficient. Thus, in addition to an increased algorithmic complexity (the search space is $\mathbb{C}^3$ instead of $\mathbb{R}^3$), more fundamental problems arise, such as the definition of the branch cuts associated to each branch point of the PGF or the definition of "physical" modes (i.e. modes that can be physically excited). To address these questions, we use the approach proposed in \cite{space_waves_1, space_waves_2, space_waves_3}: the goal is to compute the propagation of the fields along the $\dir{x}$ direction, so that branch cuts follow the steepest descent path from the branch points and only poles located on the top Riemann sheet below the real $\phi_x$ axis are considered. Using this formalism, the fields propagating along the surface can be split into two contributions: the contribution of the branch cuts (space waves) and the contributions of the poles (surface mode). While the position of the former is known in the complex plane, the position (or even existence) of the latter is not, freom there the need to find them to characterize the periodic structure.

Given the presence of complex phase shifts, the automated tracking procedure described in Section \ref{sec:ModelDisperSion} cannot be used. However, the specialized interpolation procedure and the mapping of the determinant can be straightforwardly extended to deal with complex phase shifts or frequencies. In this Section we study two different geometries taken from \cite{D17} supporting lossy modes.

\subsection{Meandered microstrip line on lossy substrate}
\begin{figure}
    \centering
    \includegraphics[width = 8.5cm]{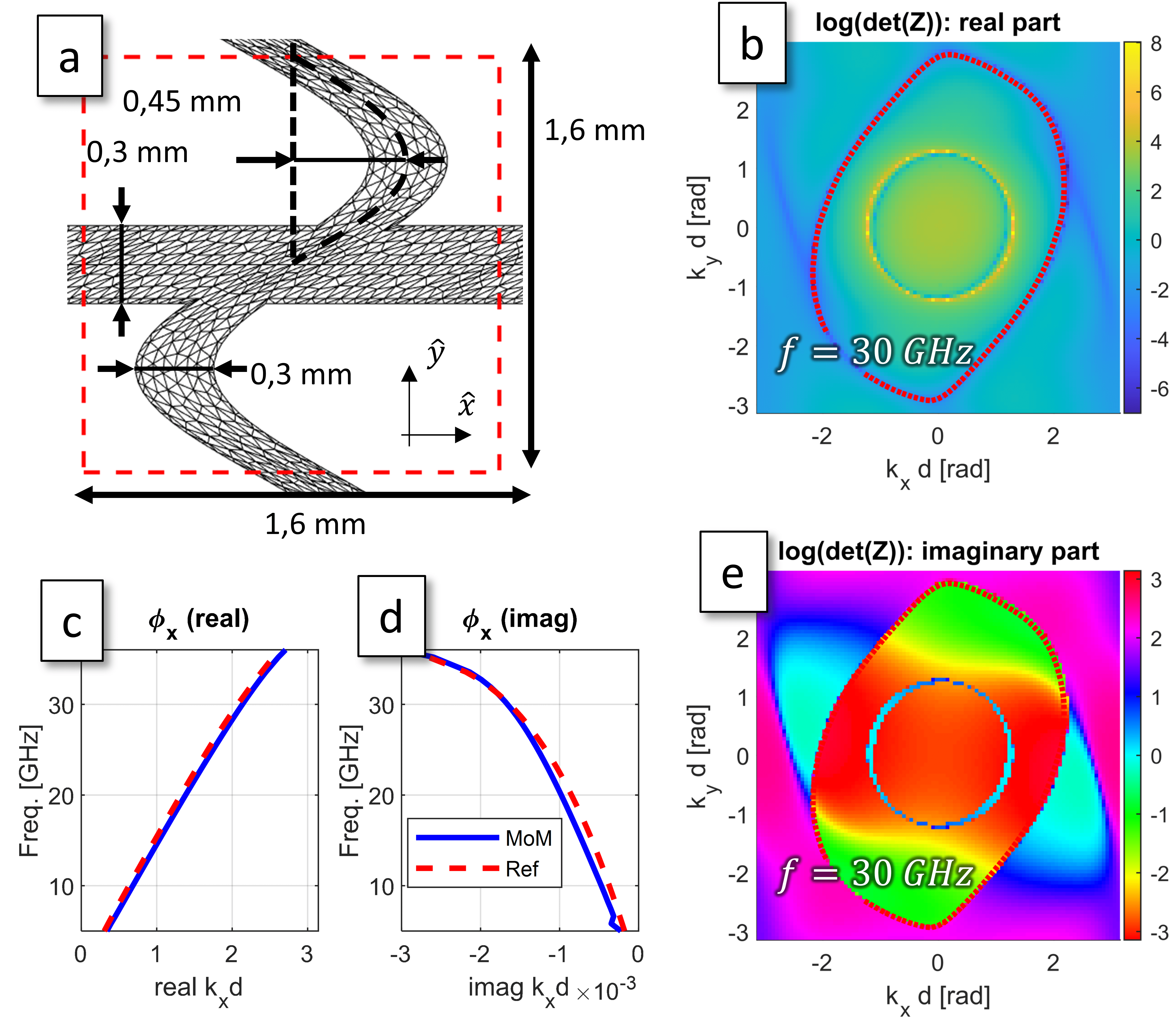}
    \caption{(a) geometry of the microstrip enclosed in the lossy parallel-plate waveguide. (b) and (d) map of the real and imaginary parts of the determinant of the periodic impedance matrix at 30 GHz. The red dashed line corresponds to the reference results from \cite{D17}. (c) and (d): Real and imaginary parts of the propagation constant of one of the guided modes in the $\dir{x}$ direction ($k_yd = 0$, $f=30$ GHz). The results of \cite{D17} extracted using the Engauge Digitizer software~\cite{engauge} have been added for comparison (red dashed line).}
    \label{fig:meander}
\end{figure}

The first geometry corresponds to a meandered microstrip enclosed in a parallel-plate waveguide. The geometry of the microstrip is displayed in Fig. \ref{fig:meander}(a). It is deposited on a 0.3 mm thick grounded dielectric layer of relative permittivity $\varepsilon_r = 3.6 (1+0.0015j)$. The air layer between the microstrip and the upper conducting plane is 0.3 mm thick. The structure support a slightly damped strongly anisotropic guided wave~\cite{D17}.

The determinant of the impedance matrix versus real phase shifts at 30 GHz is displayed in Figs. \ref{fig:meander}(b,d). Two closed curves with small value of the determinant are separated by a closed curve with a high value. The high-value curve coincides with the pole of the PGF (that is due to the fundamental mode of the parallel-plate waveguide in the absence of the printed microstripline). The two low-value curves correspond to the two guided modes supported by the waveguide in the presence of the microstrip. Since the geometry includes three different 2D-connected conducting structures, the presence of two different quasi-TEM modes was expected. It can be seen that the shape of the outer curve matches nearly perfectly the results of \cite{D17} (red dashed line). 

In addition to the two guided modes, two regions with a low value of the determinant can be seen on the left and right-hand sides of Fig. \ref{fig:meander}(b,e). These are numerical artifacts associated to the MoM itself. They are very sensitive to the mesh or the number of MBFs used, and can thus be easily discriminated from real guided modes. 

Due to the presence of losses, the determinant never perfectly vanishes for real phase shifts. To find the zeros of the determinant, one needs to look at the complex $\phi_x-\phi_y$ space. Given the increased dimensionality of the problem, we imposed $\phi_y = 0$ and mapped the determinant in the complex $\phi_x$ plane. The automated tracking of the zero in the complex plane being out of the scope of this paper, we extracted manually its position as a function of frequency. The results are visible in Figs. \ref{fig:meander}(c,d). It can be seen that the proposed method (blue line) compares very well with the prediction of \cite{D17} (red dashed line).

\subsection{Leaky waveguide}
\label{sec:cross_results}
To study the impact of radiation losses, another structure of \cite{D17} has been studied. It consists of a periodic square array of holes in a conducting sheet. The holes correspond to 0.8 mm-side squares that are repeated every 1.6 mm in the $\dir{x}$ and $\dir{y}$ directions. The sheet is separated from the ground plane by a 0.5 mm air layer. The half-space above the sheet is made of a dielectric material with a relative permittivity of 3.6. Since the permittivity of the upper half-space is higher than the permittivity of the air layer, the ``surface" modes correspond to fast waves that radiate. Given the radiation losses, the phase shifts associated with these modes is complex.

\begin{figure}
    \centering
    \includegraphics[width = 8.5cm]{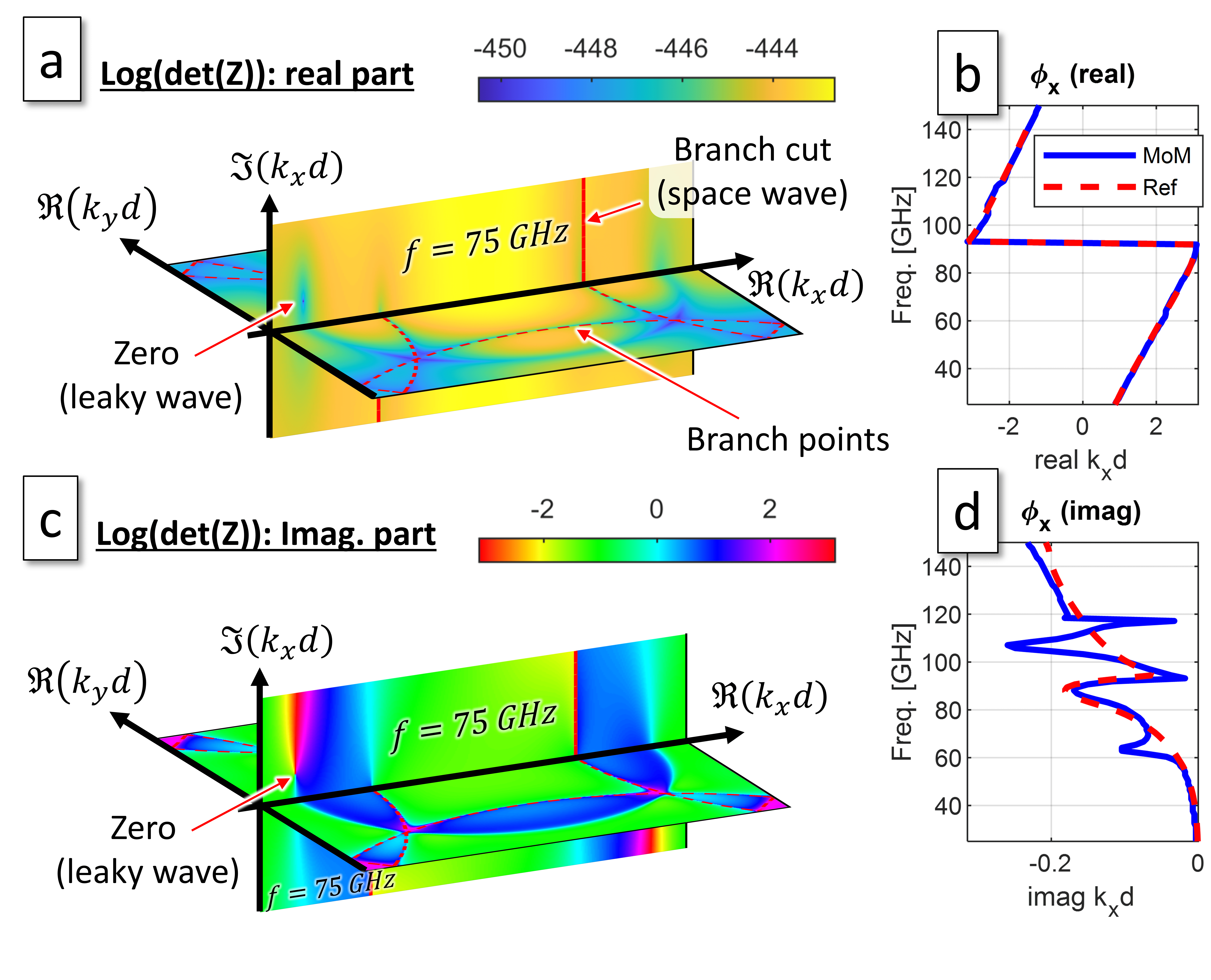}
    \caption{(a) and (c): Evolution of the real and imaginary part of the logarithm of the determinant over the first Brillouin zone for real and complex phase-shifts. The position of the branch-points (dashed red curves), branch-cuts (continuous red curve) are highlighted. (b) and (d): Real and imaginary part of the complex coordinate of the leaky wave for $\mathbb{I}\{\phi_x\} < 0$ and $\phi_y = 0$. The results of \cite{D17}, collected using~\cite{engauge}, are superimposed for comparison.}
    \label{fig:results_cross}
\end{figure}

As explained in the previous section, mapping the determinant over the 5D space corresponding to complex phase shifts and real frequency is not tractable. For this reason, we looked at the determinant of the impedance matrix for complex $\phi_x$ when $\phi_y=0$ (vertical plane in Figs. \ref{fig:results_cross}(a,c)). The zeros have been manually extracted from the determinant map at each frequency. The position of the leaky mode versus frequency is shown in Figs. \ref{fig:results_cross}(b,d) and compared with the results from \cite{D17}. 

The two curves nearly perfectly coincide for the real part of $\phi_x$. However, the imaginary parts largely differ. First, a peak of the attenuation constant is visible around 63 GHz. This peak is associated with the appearance of a grating lobe at that combination of frequency and real part of the phase shift (Wood's anomaly). To properly handle this grating lobe, one needs to account for the associated space wave (continuous spectrum corresponding to the branch cut contribution). While the periodic impedance matrix of the MoM rigorously accounts for it, the simplified transfer matrix of \cite{D17} does not. It may thus miss the possibly long-range interplay between the space wave and the leaky mode \cite{space_waves_3}. Similarly, the oscillations between 93 GHz and 118 GHz are associated with the proximity of another branch cut. At 118 GHz, the oscillating leaky mode crosses a branch cut and is replaced by a more stable leaky mode coming from the other Riemann sheet. To better illustrate the complex interactions between the zeros and the branch cuts, a video showing the evolution of the position of the zeros and branch-cuts with frequency is available in the Supplementary  Materials.

\section{Conclusion}
\label{Conclusion}
A fast eigenmode tracking technique for periodic printed structures based on the MoM has been proposed. The method is based on an efficient interpolation of the MoM impedance matrix, followed by a mapping of its determinant for various phase shifts and frequencies, and a tracking of the eigenmodes. One major advantage of the proposed interpolation method is the extremely small number of periodic impedance matrices that need to be computed explicitly. In the examples presented, only 2x2 matrices are needed to interpolate their value over the first Brillouin zone at a given frequency, and only three sampling frequencies are required to interpolate its value over more than an octave. An excellent agreement has been obtained with the results from the CST commercial software. The main advantage of the proposed method with respect to quasi-analytical methods described in the literature is the fact that it is directly built on the MoM. Therefore, it can handle arbitrary metallizations including connected ones, and very thin substrates, where semi-analytical methods may be inaccurate due to the interaction of higher order Floquet modes with the ground plane. The applicability of the proposed method to lossy structures and leaky waves has been demonstrated through several examples.

\appendices
\section{Detailed analysis of $Z^{J \rightarrow E}$}
\label{app:evol_Z}
To efficiently interpolate the impedance matrix for varying phase shifts and frequencies, we have here a deeper look at the way $Z^{J \rightarrow E}$ varies with phase shift and frequency. Looking at \eqref{eq:01}, different factors involved in each Floquet term can be highlighted~\cite{D3}. For simplicity, the Fourier transform of the BF and TF are separated into two factors: the Fourier transform of the $x-y-z$ components of the BF and TF (integral in equation \eqref{eq:03}) and the $\dir{e}$ and $\dir{m}$ unit vectors.
\begin{itemize}
\item
\textit{The $k_0$ prefactor.} It is proportional to the frequency. 
\item \textit{The $1/\kzuppq$ factor}. For high-order Floquet modes ($k_{t,pq}\gg k_0$), the derivative of this factor with frequency and phase shifts asymptotically scales as $1/k_{t,pq}^3$ and $1/k_{t,pq}^2$, respectively, i.e. variation gets smaller as the Floquet mode order increases.
\item \textit{The $\Gammate$ and $\Gammatm$ factors.} The interpolation of those factors may be complicated due to the presence of angular points, branch points and poles. The two former are due to the visibility limits in the different materials involved in the substrate. The latter is due to surface waves supported by the substrate. As a rule of thumb, the reflection coefficients exhibit smooth variation when $k_{t,pq} \gg	k_\text{max}$ with $k_\text{max}$ the wavenumber in the material with the largest refractive index. 
\item \textit{The Fourier transform of the BF and TF.} Changes in the value of the Fourier transform are induced by a change of the $\vect{k}_{pq}^\pm$ vector, which can itself be decomposed into changes in $\ktpq$ and in $\kzuppq$. For small BF and TF, variation with $\ktpq$ is quite predictable and corresponds approximately to a linear phase shift that depends on the lateral distance between the BF and TF. This phase shift is identical for all the Floquet modes and can thus be extracted easily. For large BF or TF, deviation from a linear phase shift increases. However, it should be noted that the size of the BF and TF is ultimately limited by the size of the unit cell, which also dictates the maximum variation versus $\ktpq$. Hence, in the worst-case scenario, interpolation can still provide satisfactory results, be it at the cost of a higher interpolation model. 
Concerning the variation of $\kzuppq$, we concentrate on the evolution of high-order Floquet modes. Such modes are highly evanescent, so that two cases can be highlighted. If the vertical distance between the BF and TF is short, a small change in $\kzup$ leads to a small relative variation of the Fourier transform. If the distance is large, then the product of the two Fourier transforms vanishes.
\item \textit{The $\dir{e}$ and $\dir{m}$ directions.} Rapid changes of the directions are possible when $k_{t,pq} \rightarrow 0$ or near the visibility limit. However, for high-order Floquet modes, the variation with phase shift is quite smooth. It can be noticed that $\dir{e}$ is independent from frequency, while $\dir{m}$ is inversely proportional to the frequency.
\end{itemize}

\section{Size reduction using MBFs}
\label{app:mbfs}
Model Order Reduction (MOR) techniques are based on the observation that the response of a structure to electromagnetic fields can be modelled accurately using a limited number of degrees of freedom (DoFs) \cite{D19, D7bis}, hereafter referred to as \textit{physical DoFs}. Using the MoM, those physical DoFs generally correspond to current distributions spanning over the entire structure. However, most electromagnetic solvers use BFs locally defined on a mesh to model the response of the structure. These \textit{numerical} DoFs generally poorly match the physical ones. Hence, to get accurate results, a dense mesh with many extra numerical DoFs is required. The basic idea of MOR techniques is to find, among all the numerical DoFs, those that are close to physical DoFs and discard the others. 
The MBF method consists in computing the response of the structure to a large set of different excitations \cite{D7bis}. Then, an orthonormal basis is built to approximate these responses within a chosen accuracy. In the case of the MoM, each vector of the basis corresponds to a current distribution that is close to a physical DoF of the structure.

To build the MBFs of the structure, we used a technique similar to \cite{D7bis}. The fast interpolation technique is used to get the periodic impedance matrix for several phase shifts and frequencies, (cf. Section \ref{sec:interp}). Then, the response of the structure to plane waves corresponding to low-order Floquet modes are computed. The resulting current distributions are concatenated in a large matrix and a Singular Value Decomposition (SVD) is applied to the matrix. A threshold $\theta_1$ is chosen. Modes whose singular value is smaller than the maximum singular value by a factor exceeding $\theta_1$ are discarded.

To limit the number of current distributions that should be kept in memory, intermediate SVDs are used. For any new phase shift and frequency, the newly computed current distributions are appended to the already computed singular vectors. Then, an SVD is applied to the matrix, creating a new orthonormal basis that includes the new excitations. Then, a threshold $\theta_2$ is chosen and singular vectors whose singular value (relative to the maximum singular value) is smaller than  the threshold are discarded. We used a threshold $\theta_2$ that is less restrictive than $\theta_1$ to limit the impact of the intermediate SVDs on the final results obtained. 

For the testing procedure, we used a set of Macro Testing Functions (MTFs) equal to that of the MBFs. We did not observe significant difference in the results using MTFs equal to the MBFs or equal to their complex conjugate.

The use of MBFs allows us to deal with significantly smaller matrices when searching for roots of the determinant.

\section*{Acknowledgment}
The authors would like to thank Fonds de la Recherche Scientifique - FNRS, Belgium for the funding provided for this work.
\ifCLASSOPTIONcaptionsoff
  \newpage
\fi


\begin{thebibliography}{1}

\bibitem{Numerical_methods}
C.~Craeye, X.~Radu, A.~Schuchinsky and F.~Capolino, ``Fundamentals of Method of Moments for metamaterials," \textit{in Handbook of Metamaterials, Ed. F. Capolino, Taylor and Francis}, June 2009.

\bibitem{Sheet_imped}
M.A.~Francavilla, E.~Martini, S.~Maci and G.~Vecchi, ``On the numerical simulation of metasurfaces with impedance boundary condition integral equation," \textit{IEEE Trans. Antennas Propag.}, Vol. 63, no. 5, pp. 2153-2166, 2015.

\bibitem{chapter_maci}
G.~Minatti et al., ``Metasurface Antennas," in \textit{Aperture Antennas for Millimeter and Sub-Millimeter Wave Applications}, A. Boriskin and R. Sauleau, Eds. Cham, Switzerland: Springer, 2018, ch. 9, pp. 289-333.

\bibitem{D25}
S. Enoch, G. Tayeb and B. Gralak, ``The Richness of the Dispersion Relation of Electromagnetic Bandgap Materials," \textit{IEEE Trans. Antennas Propag.}, Vol. 51, no. 10, pp. 2659-2666, 2003.

\bibitem{D27}
M. Mencagli, E. Martini, D. Gonz\'{a}lez-Ovejero and S. Maci, ``Metasurfing by Transformation Electromagnetics," \textit{IEEE Antennas Wirel. Propag. Lett.}, Vol. 13, pp. 1767-1770, 2014.

\bibitem{D29}
Q. Chen, F. Giusti, G. Valerio, F. Mesa and O. Quevedo-Teruel, ``Anisotropic glide-symmetric substrate-integrated holey measurface for a compressed ultrawideband Luneburg lens," \textit{Appl. Phys. Lett.}, Vol. 118, pp. 084102, 2021.

\bibitem{D14}
M. Dehmollaian and C. Caloz, ``General mapping between complex spatial and temporal frequencies by analytical continuation," \textit{IEEE Trans. Antennas Propag.}, Vol. 69, no. 10, pp. 6531-6545, 2021.

\bibitem{D26}
J. G. Nizer Rahmeier, V. Tiukuvaara and S. Gupta, ``Complex eigenmodes and eigenfrequencies in electromagnetics," \textit{IEEE Trans. Antennas Propag.}, Vol. 69, no. 8, pp. 4644-4656, 2021.

\bibitem{D17}
F. Giusti, Q. Chen, F. Mesa, M. Albani and O. Quevedo-Teruel, ``Efficient Bloch analysis of general periodic structures with a linearized multimodal transfer-matrix approach," \textit{IEEE Trans. Antennas Propag.}, Vol. 70, no. 7, pp. 5555-5562, July 2022.

\bibitem{D22}
A. Ali Tavallaee and J. P. Webb, ``Finite-Element modeling of evanescent modes in the stopband of periodic structures," \textit{IEEE Trans. Magn.}, Vol. 44, no. 6, pp. 1358-1361, 2008.

\bibitem{D37}
X. Xiong, J. A. Russer, W. Che, G. Shen, Y. Han and P. Russer, ``Dispersion analysis of a fishnet metamaterial based on the rotated transmission-line matrix method," \textit{IET Microw. Antennas. Propag.}, Vol. 9, no. 12, pp. 1345-1353, 2015.

\bibitem{D1bis}
D. Tihon, V. Sozio, N. A. Ozdemir, M. Albani and C. Craeye, ``Numerically stable eigenmode extraction in 3-D periodic metamaterials," \textit{IEEE Trans. Antennas Propag.}, Vol. 64, no. 7, pp. 3068-3079, 2016.

\bibitem{D19}
C. Scheiber, A. Schultschik, O. B\'{i}r\'{o} and R. Dyczij-Edlinger, ``A model order reduction method for efficient band structure calculations of photonic crystals," \textit{IEEE Trans. Magn.}, Vol. 47, no. 5, pp. 1534-1537, 2011.

\bibitem{D20}
P.-J. Chiang, C.-P. Yu and H.-C. Chang, ``Analysis of two-dimensional photonic crystals using a multidomain pseudospectral method," \textit{Phys. Rev. E}, Vol. 75, pp. 026703, 2007.

\bibitem{D24}
H. Y. D. Yang, ``Finite difference analysis of 2-D photonic crystals," \textit{IEEE Trans. Microw. Theory Tech.}, Vol. 44, no. 12, pp. 2688-2695, 1996.

\bibitem{D40}
L. Tsang and S. Tan, ``Calculation of band diagrams and low frequency dispersion relation of 2D periodic dielectric scatterers using broadband Green's function with low wavenumber extraction (BBGFL)", \textit{Opt. Express}, Vol. 24, no. 2, pp. 945-965, 2016. 

\bibitem{D15}
M. Bagheriasl, O. Quevedo-Teruel and G. Valerio, ``Bloch analysis of artificial lines and surfaces exhibiting glide symmetry," \textit{IEEE Trans. Microw. Theory Tech.}, Vol. 67, no. 7, pp. 2618-2628, 2019.

\bibitem{D16}
F. Mesa, G. Valerio, R. Rodr\'{i}guez-Berral and O. Quevedo-Teruel, ``Simulation-assisted efficient computation of the dispersion diagram of periodic structures," \textit{IEEE Antennas Propag. Mag.}, Vol. 63, no. 5, pp. 33-45, 2021.

\bibitem{D11}
P. Bienstman, H. Derudder, R. Baets, F. Olyslager and D. De Zutter, ``Analysis of cylindrical waveguide discontinuities using vectorial eigenmodes and perfectly matched layers," \textit{IEEE Trans. Microw. Theory Tech.}, Vol. 49, no. 2, pp. 349-354, 2001.

\bibitem{D23}
S. Shi, C. Chen and D. W. Prather, ``Plane-wave expansion method for calculating band structures of photonic crystal slabs with perfectly matched layers," \textit{J. Opt. Soc. Am. A}, Vol. 21, no. 9, pp. 1769-1775, 2004.


\bibitem{space_waves_1}
T. Tamir and A. A. Oliner, ``Guided complex waves. Part 1: Fields at an interface," in \textit{Proc. Inst. Electr. Eng.}, Vol. 110, no. 2, pp. 310-324, 1963.

\bibitem{space_waves_2}
F. Capolino, D. R. Jackson and D. R. Wilton, ``Fundamental properties of the field at the interface between air and a periodic artificial material excited by a line source," \textit{IEEE Trans. Antennas Propag.}, Vol. 53, no. 1, pp. 91-99, 2005.

\bibitem{space_waves_3}
F. Capolino, D. R. Jackson and D. R. Wilton, ``Field representation in periodic artificial materials excited by a source," in \textit{Theory and Phenomena of Metamaterials}, F. Capolino, Ed. Boca Raton, FL, USA: CRC Press, 2009, pp. 12.1-12.26.

\bibitem{D4bis}
X. Zheng, V. K. Kalev, N. Verellen, V. Volskiy, L. O. Herrmann, P. Van Dorp, J. J. Baumberg, G. A. E. Vandenbosch and V. V. Moschchalkov, ``Implementation of the natural mode analysis for nanotopologies using a volumetric Method of Moments (V-MoM) algorithm," \textit{IEEE Photonics J.}, Vol. 6, no. 4, pp. 4801413, 2014.

\bibitem{D5}
L. M. Delves and J. N. Lyness, ``A numerical method for locating zeros of an analytic function," \textit{Math. Comput.}, Vol. 21, no. 100, pp. 543-560, 1967.

\bibitem{D13}
D. A. Bykov and L. L. Doskolovich, ``Numerical methods for calculating poles of the scattering matrix with applications in grating theory," \textit{J. Lightwave Technol.}, Vol. 31, no. 5, pp. 793-801, 2012.

\bibitem{D36}
H. Alaeian and R. Faraji-Dana, ``A fast and accurate analysis of 2-D periodic devices using complex images Green's functions," \textit{J. Lightwave Technol.}, Vol. 27, no. 13, pp. 2216-2223, 2009.

\bibitem{D38}
H. Ameri and R. Faraji-Dana, ``Green's function analysis of electromagnetic wave propagation in photonic crystal devices using complex images techniques," \textit{J. Lightwave Technol.}, Vol. 29, no. 3, pp. 298-304, 2011.

\bibitem{D35}
S. Maci, M. Caiazzo, A. Cucini and M. Casaletti, ``A pole-zero matching method for EBG surfaces composed of a dipole FSS printed on a grounded dielectric slab," \textit{IEEE Trans. Antennas Propag.}, Vol. 53, no. 1, pp. 70-81, 2005.

\bibitem{D31}
A. M. Patel and A. Grbic, ``Modeling and analysis of printed-circuit tensor impedance surfaces," \textit{IEEE Trans. Antennas Propag.}, Vol. 61, no. 1, pp. 211-220, 2013.

\bibitem{D28}
M. Mencagli, E. Martini and S. Maci, ``Surface wave dispersion for anisotropic metasurfaces contstituted by elliptical patches," \textit{IEEE Trans. Antennas Propag.}, Vol. 63, no. 7, pp. 2992-3003, 2015.

\bibitem{D34}
S. C. Pavone, E. Martini, F. Caminita, M. Albani and S. Maci, ``Surface wave dispersion for a tunable grounded liquid crystal substrate without and with metasurface on top," \textit{IEEE Trans. Antennas Propag.}, Vol. 65, no. 7, pp. 3540-3548, 2017.

\bibitem{D32}
M. Mencagli, C. Della Giovampaola and S. Maci, ``A closed-form representation of isofrequency dispersion curves and group velocity for surface waves supported by anisotropic and spatially dispersive metasurfaces," \textit{IEEE Trans. Antennas Propag.}, Vol. 64, no. 6, pp. 2319-2327, 2016.

\bibitem{D1}
D. S. Weile, E. Michielssen and K. Gallivan, ``Reduced-order modeling of multiscreen frequency-selective surfaces using Krylov-bazed rational interpolation," \textit{IEEE Trans. Antennas Propag.}, Vol. 49, no. 5, pp. 801-813, 2001.


\bibitem{D8}
A. S. Barlevy and Y. Rahmat-Samii, ``Characterization of electromagnetic band-gaps composed of multiple periodic tripods with interconnecting vias: concept, analysis, and design," \textit{IEEE Trans. Antennas Propag.}, Vol. 49, no. 3, pp. 343-353, 2001.

\bibitem{D9}
L. Li, D. H. Werner, J. A. Bossard and T. S. Mayer, ``A model-based parameter estimation technique for wide-band interpolation of periodic moment method impedance matrices with application to genetic algorithm optimization of frequency selective surfaces," \textit{IEEE Trans. Antennas Propag.}, Vol. 54, no. 3, pp. 908-924, 2006.

\bibitem{D18}
X. Wang and D. H. Werner, ``Improved model-based parameter estimation approach for accelerated periodic Method of Moments solutions with application to the analysis of convoluted frequency selected surfaces and metamaterials," \textit{IEEE Trans. Antennas Propag.}, Vol. 58, no. 1, pp. 122-131, 2010.

\bibitem{D4}
X. Wang, D. H. Werner and J. P. Turpin, ``Application of AIM and MBPE techniques to accelerate modeling of 3-D doubly periodic structures with nonorthogonal lattices composed of bianisotropic media," \textit{IEEE Trans. Antennas Propag.}, Vol. 62, no. 8, pp. 4067-4080, 2014.

\bibitem{D6}
Z. Wang and S. V. Hum, ``A broadband model-based parameter estimation method for analyzing multilayer periodic structures," \textit{IEEE Trans. Antennas Propag.}, Vol. 69, no. 9, pp. 5771-5780, 2021.

\bibitem{D2}
D. Tihon, M. Bodehou, S. N. Jha and C. Craeye, ``Fast eigenmode mapping in printed periodic structures," \textit{Proc. of 16th European Conference on Antennas and Propagation (EuCAP)}, pp. 1-5, 2022.

\bibitem{D3}
D. Tihon, C. Craeye, N. A. Ozdemir and S. Withington, ``Interpolation of impedance matrices for varying quasi-periodic boundary conditions in 2D periodic Method of Moments," \textit{Proc. of 15th European Conference on Antennas and Propagation (EuCAP)}, pp. 1-5, 2021.

\bibitem{RAO82}
S. M. Rao, D. R. Wilton, and A. W. Glisson, “Electromagnetic scattering by surfaces of arbitrary shape,” \textit{IEEE Trans. Antennas Propag.}, Vol. 30, no. 3, pp. 409–418, 1982.

\bibitem{Reviewers_3}
M. Camacho, R. Boix and F. Median, ``NUFFT for the efficient spectral domain MoM analysis of a wide variety of multilayered periodic structures," \textit{IEEE Trans. Antennas Propag.}, Vol. 67, no. 10, pp. 6551-6563, 2019.

\bibitem{D7bis}
E. Lucente, A. Monorchio and R. Mittra, ``An iteration-free MoM approach based on excitation independent Characteristic basis functions for solving large multiscale electromagnetic scattering problems," \textit{IEEE Trans. Antennas Propag.}, Vol. 56, no. 4, pp. 999-1007, 2008.

\bibitem{MoM_metasurf_1}
S. Sandeep and S. Y. Huang, ``Simulation of circular cylindrical metasurfaces using GSTC-MoM," \textit{IEEE J. Multiscale Multiphys. Comput. Tech.}, Vol. 3, pp. 185-192, 2018.

\bibitem{MoM_metasurf_2}
J. Budhu and A. Grbic, ``Perfectly reflecting metasurface reflectarrays: Mutual coupling modeling between unique elements through homogenization," \textit{IEEE Trans. Antennas Propag.}, Vol. 69, no. 1, pp. 122-134, 2020.

\bibitem{MoM_metasurf_3}
S. Pearson and S. V. Hum, ``Optimization of electromagnetic metasurface parameters satisfying far-field criteria," \textit{IEEE Trans. Antennas Propag.}, Vol. 70, no. 5, pp. 3477-3488, 2021.

\bibitem{MoM_metasurf_4}
D. Gonzalez-Ovejero and S. Maci, ``Gaussian ring basis functions for the analysis of modulated metasurface antennas," \textit{IEEE Trans. Antennas Propag.}, Vol. 63, no. 9, pp. 3982-3993, 2015.

\bibitem{MoM_metasurf_5}
J. Cavillot, M. Bodehou and C. Craeye, ``Metasurface antennas design: Full-wave feeder modeling and far-field optimization," \textit{IEEE Trans. Antennas Propag.}, Vol. 71, no. 1, pp. 39-49, 2022.

\bibitem{FLO15}
R. Florencio, R.R. Boix, and J.A. Encinar, ``Fast and accurate MoM analysis of periodic arrays in multi-layered stacked rectangular patches with application to the design of reflectarray antennas,’’ \textit{IEEE Trans. Antennas Propag.}, Vol. 63, no. 6, pp. 2558-2571, 2015.

\bibitem{POZ84}
D.M. Pozar, D.H. Schaubert, ``Scan blindness in infinite phased arrays of printed dipoles,’’ \textit{IEEE Trans. Antennas Propag.}, Vol. 32, no. 6, pp. 602-610, 1984.

\bibitem{Jha_2014}
S. N. Jha, ``Fast spectral-domain Macro Basis Function analysis of large arrays of printed antennas," Ph.D. dissertation, ICTEAM institute, Universit\'{e} catholique de Louvain, Louvain-la-Neuve, Belgium, Aug. 2014.

\bibitem{GUE09}
Guérin N., Craeye C., Dardenne X., “Accelerated computation of the free-space Green’s function gradient of infinite phased arrays of dipoles,” \textit{IEEE Trans. Antennas Propag.}, Vol. 57, no. 9, pp. 3430-3434, 2009.

\bibitem{Levin_T}
S.~Singh and R.~Singh, ``On the use of Levin's T-transform in accelerating the summation of series representing the free-space periodic Green's functions," \textit{IEEE Trans. Mircrow. Theory Tech.}, Vol. 41, no. 5, pp. 884-886, 1993.

\bibitem{CRA04}
Craeye C., Tijhuis A.G., Schaubert D.H., ``An efficient MoM formulation for finite-by-infinite arrays of two-dimensional antennas arranged in a three-dimensional structure," \textit{IEEE Trans. Antennas Propag.}, vol. 52, no. 1, pp. 271-282, 2004.

\bibitem{WIL84}
D. R. Wilton, S. M. Rao, A. W. Glisson, D. H. Schaubert, O. M. Al-Bundak, and C. M. Butler, “Potential integrals for uniform and linear source distributions on polygonal and polyhedral domains,” \textit{IEEE Trans. Antennas Propag.}, Vol. 32, no. 3, pp. 276–281, 1984.

\bibitem{Reviewers_1}
S. Monni, G. Gerini, A. Neto and A. G. Tijhuis, ``Multimode equivalent networks for the design and analysis of frequency selective surfaces," \textit{IEEE Trans. Antennas Propag.}, Vol. 55, no. 10, pp. 2824-2835, 2007.


\bibitem{Reviewers_2}
R. Rodr\'iguez-Berral, F. Mesa and F. Medina, ``Analytical multimodal network approach for 2-D arrays of planar patches/apertures embedded in a layered medium," \textit{IEEE Trans. Antennas Propag.}, Vol. 63, no. 5, pp. 1969-1984, 2015.

\bibitem{MTSforspace}
G.~Minatti, M.~Faenzi, E.~Martini, F.~Caminita, P.~De Vita, D.~Gonz\'{a}lez-Ovejero, M.~Sabbadini and S.~Maci, ``Modulated metasurface antennas for space: synthesis, analysis and realizations," \textit{IEEE Trans. Antennas Propag.}, vol. 63, pp. 1288-1300, Apr. 2015.

\bibitem{inverion_SW_antenna}
M.~Bodehou, K.~Alkhalifeh, S.~N.~Jha, and C.~Craeye, ``Direct numerical inversion methods for the design of surface-wave based metasurface antennas: fundamentals, realization, and perspectives," \textit{IEEE Antennas Propag. Mag.}, vol. 64, no. 4, pp. 24-36, June 2022.

\bibitem{CSTref}
Dassault Systemes, CST Studio Suite, Electromagnetics Simulation Software, https://www.3ds.com/products-services/simulia/products/cst-studio-suite/solvers/

\bibitem{engauge}
M. Mitchell, B. Muftakhidinov, T. Winchen et al., "Engauge Digitizer Software." Available at http://markummitchell.github.io/engauge-digitizer

\end{thebibliography}
\end{document}